\pdfoutput=1
\documentclass[sigplan,nonacm]{acmart}
 
\usepackage{amsmath}
\usepackage{array}
\usepackage{booktabs}
\usepackage{graphicx}
\usepackage{microtype}
\usepackage{multirow}
\usepackage{placeins}
\usepackage{subcaption}
\usepackage{tabularx}
\usepackage{tikz}
\usepackage{xcolor}
\usepackage{xspace}
\usepackage{algorithm}
\usepackage{algpseudocode}

\usetikzlibrary{arrows.meta,fit,positioning,shapes.geometric}

\newcolumntype{Y}{>{\raggedright\arraybackslash}X}
\newcolumntype{C}{>{\centering\arraybackslash}X}

\newcommand{\system}{Koi\xspace}

\newcommand{\placeholder}[1]{\textcolor{red}{\textbf{[#1]}}}

\definecolor{leafgreen}{HTML}{519A66}
\definecolor{cherryblossom}{HTML}{F891BB}

\newcommand{\gmnew}[1]{{\color{leafgreen}{#1}}}

\renewcommand{\shortauthors}{Sidhu et al.}

\author{
Mankeerat Sidhu\textsuperscript{1,*},
~Gangmuk Lim\textsuperscript{2,*},
~Hetarth Chopra\textsuperscript{1},
~Maximillian Eisenberg\textsuperscript{1}, \\
~Quan Hao Ng\textsuperscript{1},
~Aryan Gosaliya\textsuperscript{1},
~Shu Liu\textsuperscript{3},
~Luka Govedi\v{c}\textsuperscript{4}, \\
~Jiaxin Shan\textsuperscript{5},
~Hongkuan Zhou\textsuperscript{6},
~Vikram Sharma Mailthody\textsuperscript{6},
~P. Brighten Godfrey\textsuperscript{2}, \\
~Ion Stoica\textsuperscript{3}
\\[1em]
\textsuperscript{1}Tandemn \quad
\textsuperscript{2}University of Illinois Urbana-Champaign \quad
\textsuperscript{3}UC Berkeley \\
\textsuperscript{4}New York University \quad
\textsuperscript{5}ByteDance \quad
\textsuperscript{6}NVIDIA
\\[0.5em]
\texttt{\{mankeerat, hetarth, max, quan, aryan\}@tandemn.com}, \texttt{\{gangmuk2, pbg\}@illinois.edu} \\
\texttt{\{lshu, istoica\}@berkeley.edu}, \texttt{lg3769@nyu.edu} \\
\texttt{jiaxin.shan@bytedance.com}, \texttt{\{hongkuanz, vmailthody\}@nvidia.com}
}
\affiliation{\institution{}\country{USA}}

\begin{document}

\title{\system: Evidence-Guided, Self-Calibrating Planning for Multi-Cloud, Heterogeneous LLM Inference Fleets}

\begin{abstract}

Deploying a large language model (LLM) requires choosing the right accelerator, parallelization strategy, engine configuration, and replica count amongst $\sim$140 other variables across clouds, regions, and markets. These decisions are intertwined, and the best cluster-wide plan continually changes as jobs contend for shared capacity on shifting workloads. 

We present \system, a self-calibrating cluster planner for heterogeneous, multi-cloud fleets. \system uses a Recursive Language Model (RLM) hierarchy to propose hypotheses and reason over deployment while deterministic components predict serving behavior, score candidates, and control placement. It represents inference behavior as a fixed causal graph linking decisions (e.g., accelerator, parallelism), runtime mediators (e.g., KV-cache pressure, communication overhead), and outcomes (e.g., latency, throughput). Each deployment strengthens or weakens scoped causal relationships, building a comprehensive playbook for inference that the planner employs to guide future decisions. Job-wise configurations are composed into cluster-wide plans and evaluated against serving objectives, resource constraints, SLO risks, and cost.

Across benchmarks, \system achieves $1.18\times$ greater goodput than the strongest baseline while reducing \$/token by 20\% in a real heterogeneous cluster. Ablations isolate the contribution of \system's architecture: a pure language model attains a composite score of 35.3, the same model equipped with \system's deterministic tools reaches 51.6, and the full \system planner reaches 70.8.

\end{abstract}



\maketitle

\renewcommand{\thefootnote}{*}
\footnotetext{Co-first authors.}

\pagestyle{plain}

\section{Introduction}
\label{sec:introduction}


An LLM inference provider serves a changing mix of jobs on a shared accelerator fleet that spans clouds, regions, markets (reserved, on-demand, or spot), and hardware generations. Users specify each job's model, workload profile (input/output lengths and arrival process), class (online or batch), and serving target (a latency bound for online jobs or a completion deadline for batch jobs). The provider decides how every job runs, including the environment and accelerator, the parallelism and engine configuration, and the number of replicas, and must revise these decisions as jobs arrive and complete and as available capacity changes.

\begin{figure}[H]
  \centering
  \includegraphics[
    width=0.58\columnwidth,
    keepaspectratio
  ]{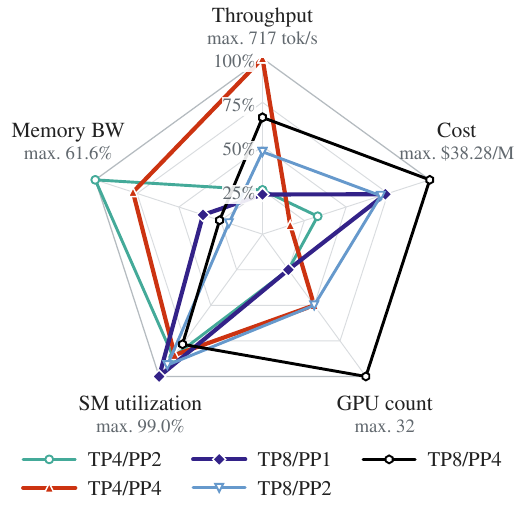}
    \caption{\textbf{Parallelism configurations expose conflicting performance, cost, and utilization trade-offs.} It used Llama~3 70B on AWS A10G GPUs with 2,048/512 input/output tokens. Each axis is normalized independently.}
  \label{fig:motivation-placement-radar}
\end{figure}

\begin{figure}[t]
  \centering
  \includegraphics[
    width=\columnwidth,
    keepaspectratio
  ]{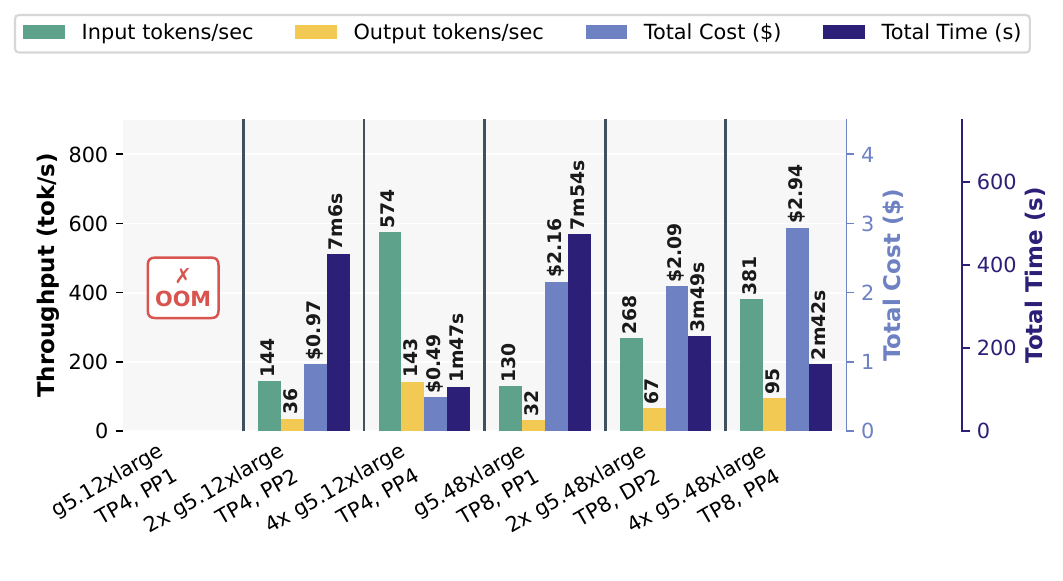}
  \caption{\textbf{Parallelism substantially changes inference performance and cost.} Done on AWS A10G instances with Llama~3 70B model. The workload has 2,048 input and 512 output tokens per request.}
\label{fig:motivation-parallelism}
\end{figure}

Even for a single job, this choice is consequential. For a profiled 70B dense model on one accelerator family, five feasible tensor- and pipeline-parallel configurations span $4.4\times$ in throughput and $7.5\times$ in cost per token and trade off performance, cost, and utilization differently (Figures~\ref{fig:motivation-placement-radar} and~\ref{fig:motivation-parallelism}). TP4/PP2 across two four-GPU instances serves the workload at 180~tok/s for \$0.97, whereas TP8/PP1 on one eight-GPU instance reaches only 162~tok/s for \$2.16. The best configuration also depends on workload shape, because prefill-heavy traffic stresses compute while decode-heavy traffic stresses memory bandwidth and KV-cache capacity. Neither the workload nor the serving stack is stationary. A job's realized input/output distribution drifts, and engine and library updates change throughput for the same deployment, so a configuration profiled once does not stay optimal and the planner must keep recalibrating from what it observes.

These choices are also coupled across jobs that compete for the same finite accelerators, quota, and budget. Most schedulers decide one job at a time, giving each arriving job its best-ranked feasible placement, so scarce capacity goes to whichever job arrives first. This is suboptimal whenever an earlier job has a workable second choice that a later job lacks. For example, an online chat job and a batch evaluation job both prefer the cluster's only free H100 node, but only the batch job can meet its target on A100s. If the batch job arrives first, a greedy scheduler gives it the H100, and the online job must miss its SLO or wait. Even a correct per-job ranking cannot prevent this, because avoiding it requires weighing the opportunity cost of giving a scarce device to one job rather than another, which only a cluster-wide decision can do.

Existing systems address parts of this problem. Serving engines optimize request execution within a given deployment~\cite{vllm,orca,sarathi}. Multi-cloud orchestrators place workloads across clouds and markets but leave the accelerator type and configuration to the user~\cite{skypilot,skyserve}. Configuration optimizers and heterogeneous serving planners choose parallelism, placement, or GPU types for a single model or a fixed set of models~\cite{xu2026aiconfigurator,alpaserve,melange,helix}. Request schedulers decide how requests are queued and served on existing deployments~\cite{qlm,qoserve}, and cluster schedulers allocate resources among competing jobs~\cite{tiresias,pollux,cilantro}, but neither chooses the deployment configuration of an inference job. No existing system jointly chooses deployment configurations and allocates shared capacity for all online and batch inference jobs across a heterogeneous, multi-cloud fleet, while recalibrating those choices as observed behavior drifts.

Joint planning is hard for two reasons. First, the space is vast. Each job's deployment sets about 140 configuration knobs (Appendix~\ref{app:knobs}), including accelerator, parallelism, engine, quantization, batching, and KV-cache settings, which yields at least $2^{140}$ configurations per job. A cluster plan chooses one deployment for every job under shared capacity, so the number of candidate plans grows exponentially with the number of jobs and cannot be enumerated. Second, the planner must predict the performance of never-run deployments for every job and candidate. Analytical models and profile-based simulators~\cite{vidur,xu2026aiconfigurator} rest on fixed assumptions about memory, communication, and engine behavior that break as workloads, hardware, and software change. In our measurements, even a roofline model calibrated with measured peak compute and memory bandwidth errs by up to $4.7\times$. Language models encode broad systems knowledge but not calibrated, deployment-specific quantities. Two frontier models, Claude Opus~4.8 and GPT-5.6 Sol, overpredict throughput by $4\times$ and $9\times$. A planner therefore cannot trust predictions unless it corrects them from observed behavior.

Search-based methods instead deploy candidate configurations, observe their end-to-end performance, and use it to choose the next candidates. Bayesian optimization fits a statistical model to these observations~\cite{cherrypick,scoot}, and LLM-guided evolutionary search asks a language model to propose new candidates from high-scoring ones~\cite{funsearch,alphaevolve,sharma2025openevolve,liu2026evox}. These methods have two fundamental limitations. First, because they learn only from end-to-end outcomes, they cannot tell why a candidate performed as it did, for example whether low throughput came from pipeline bubbles or KV-cache pressure, and therefore which fix applies or which other jobs the evidence should inform. Second, every trial is a live deployment that occupies capacity other jobs need and can breach a running job's SLO. Evolutionary search needs many such trials, and Bayesian optimization restarts for each new workload. LLM-guided evolutionary search pairs a language model with measurement only loosely, because the model sees only a scalar score and measurements never confirm or refute its reasoning. A tighter pairing is difficult because the model's knowledge is qualitative while telemetry is numeric.

\system pairs them tightly through a fixed causal graph on which both can operate. The graph links deployment decisions (e.g., accelerator, parallelism), runtime mediators (e.g., KV-cache pressure, communication overhead), and serving outcomes (e.g., latency, throughput). Scoped \emph{mechanisms} compose its edges into testable hypotheses, which a language model can propose and telemetry can support or contradict. For example, one mechanism states that raising pipeline parallelism lowers throughput through pipeline bubbles for a given model and hardware. Each edge and mechanism carries a confidence and an uncertainty updated from runtime evidence, so confident relationships guide exploitation and uncertain ones identify deployments worth testing. \system runs no separate trials, since every deployment both serves its job and tests the relationships it exercises, and the plan score values both. Bayesian optimization also updates a model after each trial, but that model maps configurations to end-to-end outcomes for a single workload. \system instead updates its beliefs about mechanisms, so evidence from one job informs every job within the mechanism's scope.

\system turns this evidence into a cluster plan by alternating between language-model proposals and deterministic evaluation. For each job, the language model proposes candidate deployments from the causal graph, the cluster snapshot, the job's workload, and the matching mechanisms. The graph lets it focus on the decisions that affect the job's targets, so \system evaluates a few candidates per job instead of enumerating the space. Deterministic components predict each candidate's behavior with a profile-based performance model whose error is tracked against observed outcomes, score it against the job's weighted objectives, and compose candidates into a cluster-wide plan. The plan score balances serving value, the value of testing uncertain relationships, SLO risk under prediction error, and reconfiguration cost, subject to shared capacity. Because candidates for all jobs compete within one plan, this score exposes the opportunity costs that per-job optimization misses. Only deployment evidence updates causal confidence, and the language model only proposes and cannot override measured evidence or cluster constraints.

We evaluate \system on a real 104-GPU cluster spanning three regions and on simulated 400- and 800-GPU clusters, all heterogeneous, against nine baselines including the approaches above (three on the real cluster). Against the strongest baseline, \system delivers $1.18\times$ the goodput at 20\% lower \$/token on the real cluster and 45\% more goodput at 31\% lower \$/token in the 3-hour simulation. Ablations show that these gains come mainly from the structure around the language model rather than from the model itself.

This paper makes the following three contributions.
\begin{itemize}
    \item We design and implement \system, a provider-side control plane that self-calibrates the cluster-wide deployment plan for mixed online and batch jobs across heterogeneous clouds, and release it as open source\footnote{https://github.com/Tandemn-Labs/tandemn-koi}.

    \item We develop an evidence-guided agentic planning algorithm around a causal DAG that links deployment decisions, runtime mediators, and outcomes. Our RLM~\cite{zhang2026recursivelanguagemodels} hierarchy proposes candidates and scoped hypotheses, deterministic evaluation selects the cluster plan, and telemetry updates the evidence that guides subsequent intervals.

    \item We evaluate \system on heterogeneous benchmarks spanning ten model architectures and nine baselines, where it improves goodput by up to 45\% and lowers \$/token by up to 31\% while admitting 96\% of online and 100\% of batch jobs over 12 hours.
\end{itemize}

\

\section{Background}
\label{sec:background}
 
\subsection{Where deployment planning sits}
 
LLM serving spans three control layers. Serving engines admit and batch requests~\cite{orca,vllm,sarathi}, routers distribute requests among running replicas~\cite{qlm,qoserve}, and a deployment controller chooses the hardware, parallelism, engine configuration, replica count, and placement within which those layers operate. \system composes with serving engines and routers while operating at the third layer, where it plans and controls deployment.
The separation is operational as well as temporal. Engine and router decisions
are made per request and are reversible within milliseconds, whereas a
deployment decision commits model weights to devices. Changing one requires
loading weights, warming caches, and draining in-flight work, which bounds how
many configurations a controller can try in a given period and requires damping
against churn.
 
\subsection{What determines deployment performance}
\label{sec:bg-factors}

The deployment objective is not to maximize raw throughput or GPU utilization.
Rather, the controller should select the least-cost configuration that satisfies
the workload's service objectives: online workloads must meet TTFT, TPOT, and
throughput targets, while batch workloads must complete by their deadlines.
Operating closer to saturation can reduce cost, but it also reduces queueing
headroom and can sharply degrade tail latency under bursty demand.

The lowest-cost feasible configuration is workload-dependent because different
workload characteristics expose different hardware bottlenecks. Input-heavy
requests make prefill compute the limiting resource and first manifest
as elevated TTFT. Long-context and output-heavy requests increase KV-cache
residency and HBM-bandwidth pressure, which can degrade TPOT and reduce the
capacity available to concurrent requests. Arrival rate and burstiness determine
the replication and queueing headroom required to preserve these objectives.
Consequently, a model identifier and mean request rate alone are insufficient to
determine the appropriate GPU type, model-parallel configuration, or replica
count.

Configuration choices affect end-to-end objectives indirectly through runtime
mechanisms. For example, pipeline parallelism can reduce per-GPU memory pressure
but introduce pipeline bubbles and stage imbalance; similarly degraded throughput
can instead arise from KV-cache pressure or interconnect contention. These
mechanisms require different corrective actions. Thus, a controller must use
runtime evidence to identify the binding bottleneck and select the least-cost
feasible intervention, rather than relying on repeated configuration trials
guided only by endpoint metrics. This motivates \system's evidence-conditioned
control loop.

\section{Related Work}
\label{sec:related_works}

\subsection{Deployment configuration and placement}

Per-job optimizers evaluate alternative serving configurations against workload and SLO objectives. AIConfigurator predicts configurations across parallelism strategies, serving modes, and engine-specific settings, then ranks them by derived latency and throughput~\cite{xu2026aiconfigurator}.

Other systems extend deployment decisions to concurrent jobs. AlpaServe places and parallelizes multiple models to improve serving efficiency under bursty workloads~\cite{alpaserve}. Prism improves multi-model serving through demand-aware cross-model GPU-memory sharing, model placement, and request arbitration~\cite{prism}. Heterogeneous serving planners optimize a single model across mixed hardware: M{\'e}lange computes a minimum-cost heterogeneous GPU allocation for a given workload and latency target~\cite{melange}, while Helix jointly optimizes model placement and request scheduling over heterogeneous GPUs and network links~\cite{helix}. Cloud systems operate at the infrastructure layer: SkyPilot places workloads across clouds~\cite{skypilot}, while SkyServe distributes spot and on-demand replicas across regions and clouds~\cite{skyserve}. 

\subsection{Feedback-driven and causal optimization}

Measurement-driven optimizers refine decisions from observed outcomes. CherryPick uses Bayesian optimization to test cloud configurations and select subsequent trials from observed execution~\cite{cherrypick}. SCOOT applies Bayesian optimization to inference-engine parameters, using learned constraints to avoid invalid configurations while optimizing SLOs~\cite{scoot}. Cluster schedulers also adapt from runtime feedback: Pollux models how resource allocation and batch size affect training goodput~\cite{pollux}; Cilantro learns online resource-to-performance mappings under application-level objectives~\cite{cilantro}; and Tiresias schedules distributed training jobs from observations and prior knowledge of job duration~\cite{tiresias}.

Causal methods instead use structure to determine which interventions or relationships should generalize. Invariant causal prediction identifies predictors whose relationship with a target remains stable across environments~\cite{peters2016}, while causal Bayesian optimization selects interventions on a supplied causal graph to optimize a target with fewer experiments~\cite{aglietti2020cbo}.

\subsection{LLM-guided search}

FunSearch uses an LLM to modify a target function and an evaluator to score each candidate~\cite{funsearch}. AlphaEvolve extends this evolutionary search to larger programs and codebases, retaining evaluated programs to guide subsequent proposals~\cite{alphaevolve}. OpenEvolve maintains a diverse set of high-scoring programs and uses them to guide subsequent mutations~\cite{sharma2025openevolve}. EvoX further co-evolves candidate solutions and the search strategy that selects and varies them, allowing the exploration policy itself to adapt during optimization~\cite{liu2026evox}.
These systems evolve executable artifacts by repeatedly proposing and scoring candidate programs.

\section{Problem Formulation}
\label{sec:problem-formulation}

At each $n$-minute tick, \system selects a deployment configuration for every job in the cluster. At tick $t$, the cluster contains active jobs $\mathcal{A}_t$ and pending jobs $\mathcal{Q}_t$. Each job $i$ has a workload profile $W_i$ and target $S_i$: a latency bound for online jobs or a completion deadline for batch jobs. Cluster capacity is represented by a resource map $R_t$ over environments $e$.

A deployment is organized into the following hierarchy:
\textbf{Chain.}
A chain $c=(e,x)$ is one model copy deployed in environment $e$ under one deployment configuration.

\noindent\textbf{Rank.}
A rank $(c,n)$ consists of $n$ homogeneous replicas of the same chain,
where $n$ is its data-parallel size.

\noindent\textbf{Ladder.}
A ladder
$L_i=\{(c_i^1,n_i^1),\dots,(c_i^k,n_i^k)\}$
is the set of ranks assigned to job $i$.

\noindent\textbf{Plan.}
A cluster plan
$P_t=\{L_i\}_{i\in\mathcal{A}_t\cup\mathcal{Q}_t}$
assigns one ladder to every active and pending job in the cluster.
For an active job, $L_i=L_i^{(t-1)}$ retains its current deployment; for a
pending job, $L_i=\varnothing$ defers admission.

Each ladder $L_i$ produces five outcomes (\$/tok, p99 TTFT, p99 TPOT,
throughput, and SLO margin) collectively denoted $y_i(L_i)\in\mathbb{R}^5$.
The first three are minimized and the latter two maximized. 

Each job $i$ specifies objective weights
$w_i=(w_{i1},\ldots,w_{i5})$, with $w_{ij}\ge 0$ and
$\sum_j w_{ij}=1$, over $y_i(L_i)$. These weights encode the user's relative priorities for that
job's serving objectives. \system reduces these five objectives to a single scalar using an augmented Tchebycheff scalarization~\cite{miettinen1999nonlinear}. Let $z_i^\star$ be a reference point for job $i$ that weakly dominates every outcome the job can achieve, maintained per job and updated from realized outcomes (\S\ref{sec:koi-slow-loop}), and let
$\sigma_{ij}$ be a normalizing constant per objective. The normalized gap along
objective $j$ is $g_{ij}(L_i) = \bigl(y_{ij}(L_i) - z_{ij}^\star\bigr)/\sigma_{ij}$ if $j$ is
minimized and $\bigl(z_{ij}^\star - y_{ij}(L_i)\bigr)/\sigma_{ij}$ if $j$ is
maximized, and the ladder's exploitation score is
{
\begin{equation}
J_i(L_i)=-\left[
\max\nolimits_{j\in[1,5]} w_{ij}g_{ij}(L_i)
+\rho\sum\nolimits_{j=1}^{5} w_{ij}g_{ij}(L_i)
\right].
\label{eq:tchebycheff}
\end{equation}
}
where $\rho > 0$ is an augmentation constant. The max term ensures that no
single objective is neglected on non-convex Pareto frontiers; the weighted sum
breaks ties among candidates with the same maximum gap. $J_i(L_i) \le 0$
always, and $J_i(L_i) = 0$ only when every objective the job weights attains its
reference.

Let $\mathrm{Switch}(L_i^{(t-1)}, L_i)$ denote the reconfiguration cost defined in
\S\ref{sec:switch_cost}, with $\mathrm{Switch}(L,L)=0$ for an unchanged ladder.
\system selects a cluster-wide plan by maximizing aggregate ladder value minus
this cost:
{
\begin{equation}
\max\nolimits_{\{L_i\}}
\sum\nolimits_{i \in \mathcal{A}_t \cup \mathcal{Q}_t} J_i(L_i)
-\lambda_t
\sum\nolimits_{i \in \mathcal{A}_t}
\mathrm{Switch}\!\left(L_i^{(t-1)}, L_i\right)
\label{eq:tick_program}
\end{equation}
}
subject to

\noindent\textbf{C1} \textbf{Resource feasibility.}
The chains requested across all ladders must fit within the available
capacity under $R_t$.

\noindent\textbf{C2} \textbf{Physical feasibility.}
Each chain must fit available VRAM; TP must divide the attention and KV
head counts; PP must partition the model by layers.

\noindent\textbf{C3} \textbf{SLO feasibility.}
Each ladder must satisfy $S_i$ under prediction uncertainty, with
$P(\text{breach}) \le \tau$.

\noindent\textbf{C4} \textbf{Swap budget.}
At most $B_t$ active jobs may change ladders during one tick:
$|\{i \in \mathcal{A}_t : L_i \ne L_i^{(t-1)}\}| \le B_t$.

\noindent\textbf{C5} \textbf{Job policy.}
Each job must remain within its capacity quota, and higher-priority jobs
must take precedence over lower-priority jobs competing for the same resources.
    
A pending job is deferred when no ladder satisfies \textbf{C1--C3}, or when
job priority, quota, or resource contention favors reserving capacity for
another job.
 
Equation~\eqref{eq:tick_program} defines the decision \system would make if $y_i(L_i)$ were known for every candidate ladder. In practice, $L_i \mapsto y_i(L_i)$ is unknown and changes over time. Because evaluating a ladder requires deployment, consuming cluster capacity and risking SLO violations, \system instead estimates $y_i$ from available evidence, deploys, observes the outcome, and updates its estimate for the subsequent tick.

\section{\system}
\label{sec:design}
 
\subsection{Overview}
\label{sec:overview}
 
\begin{figure*}[t]
\centering
\setlength{\fboxrule}{0.4pt} 
\setlength{\fboxsep}{0pt}    
\fbox{%
    \includegraphics[
        width=\dimexpr\textwidth-2\fboxsep-2\fboxrule\relax,
        keepaspectratio
    ]{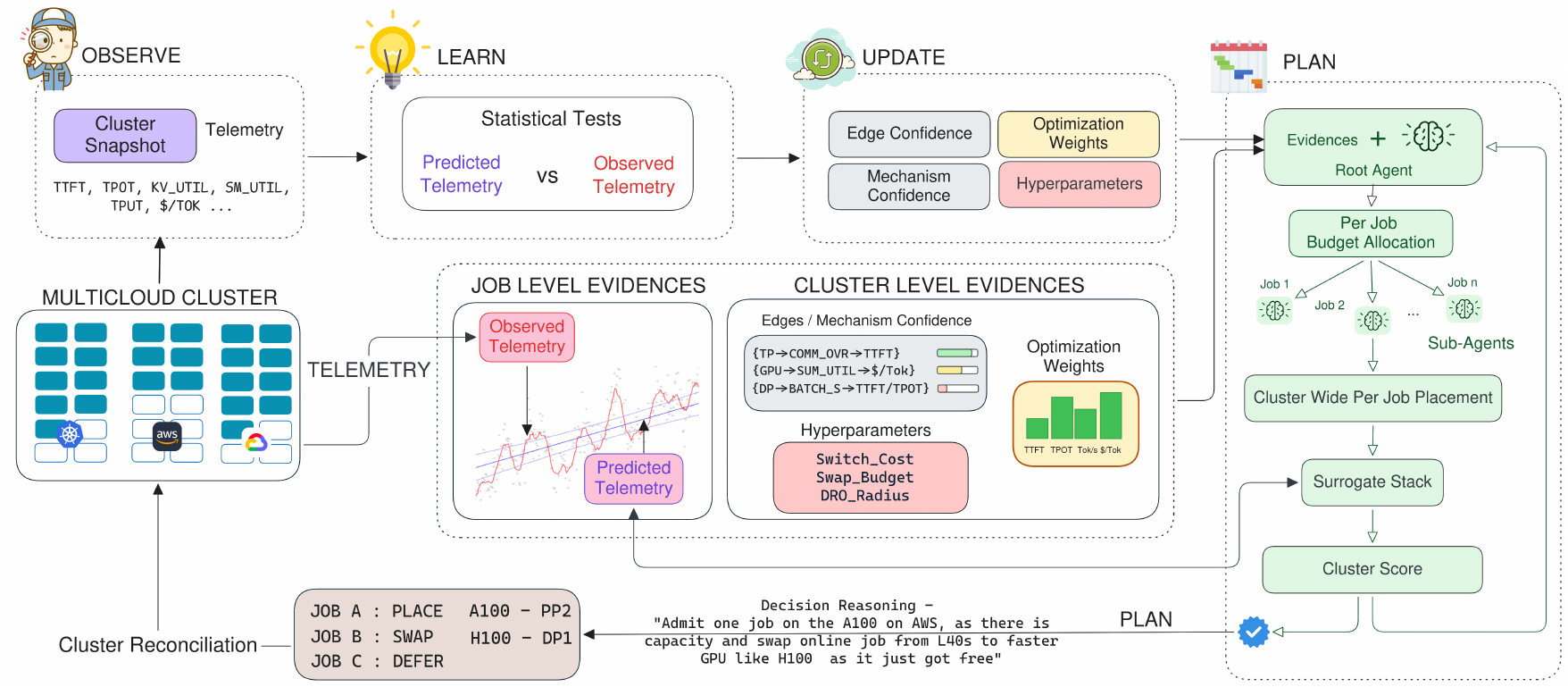}%
}
\caption{\textbf{End-to-end \system architecture.} Each planning tick snapshots cluster state, observes runtime telemetry, updates causal confidence, proposes deployment candidates, and deterministically evaluates and selects a cluster-wide plan.}
\Description{Full end-to-end \system system architecture.}
\label{fig:koi-architecture}
\end{figure*}

Each tick, \system executes an eight-stage closed loop that updates evidence from the cluster and searches for a better cluster-wide plan (Figure~\ref{fig:koi-architecture}). States S0--S3 refresh the planner state by snapshotting jobs, placements, and capacity (S0), collecting per-rank telemetry (S1), comparing observations against stored predictions (S2), and updating the evidence state (S3). Planning begins at S4, where the agent proposes a cluster-wide action for each job (place, reconfigure, retain, or defer). Deterministic evaluation at S5 scores the proposal against job-specific performance, expected evidence, SLO risk, and reconfiguration cost, and validates constraints \textbf{C1--C5}. Invalid or non-improving plans return to S4 with scores and diagnostics. Once the search budget is exhausted, \system commits the best feasible cluster-wide plan (S6) and closes the loop (S7). This design relies on two assumptions: near-optimal proposal coverage (Assumption~\ref{assump:proposal-coverage}) and convergence under stationary conditions (Assumption~\ref{assump:convergence}).
 
 

\subsection{Mechanisms and Predictions}
\label{sec:mechanisms}
 
To make deployment behavior testable and transferable, \system represents inference as a directed acyclic graph (DAG), $X \rightarrow V \rightarrow Y$, where $X$ contains planner-controlled decisions (hardware, engine, parallelism, quantization, batching, routing), $V$ contains telemetry-derived runtime mediators (KV-cache utilization, pipeline bubbles, communication latency, memory pressure, SM utilization), and $Y$ contains the serving outcomes defined in \S\ref{sec:problem-formulation}. Only $X\rightarrow V$ and $V\rightarrow Y$ edges are permitted, ensuring every modeled effect passes through telemetry; the registered edge set $E \subseteq (X \times V) \cup (V \times Y)$ is fixed at initialization.
 
Each edge $e\in E$ maintains a belief
$p_e\sim\mathrm{Beta}(\alpha_e,\beta_e)$, where $\alpha_e$ and $\beta_e$
accumulate supporting and contradictory evidence, and
$c_e=\alpha_e/(\alpha_e+\beta_e)$ is its confidence. Because the same confidence
can reflect different amounts of evidence (supporting and contradictory deployment behavior), \system also tracks normalized
posterior uncertainty
\begin{equation*}
U(\alpha_e,\beta_e)=
\frac{12\alpha_e\beta_e}
{(\alpha_e+\beta_e)^2(\alpha_e+\beta_e+1)},
\qquad U(1,1)=1.
\end{equation*}
Exploration (\S\ref{sec:exploration}) prioritizes edges with high $U$, where
additional deployment evidence is most informative.

Edges compose into scoped hypotheses called \emph{mechanisms}. A mechanism
$m=(E_m,S_m,\eta_m,B_m)$ pairs a connected subgraph $E_m\subseteq E$ with an
applicability scope $S_m$ over models, hardware, workloads, and conditions, a
natural-language statement $\eta_m$, and Beta state $B_m$. The LLM proposes
mechanisms, which are admitted only if they form a valid, non-duplicate
$X\!\rightarrow\!V\!\rightarrow\!Y$ path over registered edges and are initialized
with $B_m=\mathrm{Beta}(1,1)$.

Both $\widehat{V}$ and $\widehat{Y}$, the predicted mediator and outcome values, come from Koi's planner-internal deterministic surrogate. The surrogate builds on AIConfigurator \cite{xu2026aiconfigurator}. Given a job's workload profile and one candidate ladder, it composes operator-level latency profiles with a replay simulator to predict TTFT, TPOT, and
throughput point estimates. The remaining mediators and outcomes are derived from those
predictions, the job's targets, and the allocation price carried by $R_t$. When a model has no measured profile, it substitutes the nearest profiled architecture on that hardware and rescales by its measured
prefill and decode speed ratios. This surrogate is used during planning to score candidates and is distinct from LLMServingSim 2.0, which is used only as the execution environment for the simulated evaluation \ref{sec:evaluation}.

The surrogate is an approximation, so \system carries prediction uncertainty into
planning and rejects ladders with $P(\text{breach})>\tau$ under \textbf{C3}.
A deployed ladder tests every mechanism whose scope it satisfies by comparing the predicted mediator and outcome trajectories, $\widehat{V}$ and $\widehat{Y}$, against the observed $V_{\mathrm{obs}}$ and $Y_{\mathrm{obs}}$. For either trajectory $O\in\{V,Y\}$, \system defines the residual at observation $k$ as $r_k=O_{\mathrm{obs},k}-\widehat{O}_k$ and accumulates it in a two-sided \emph{cumulative sum} (CUSUM)~\cite{page1954continuous} detector:

\begin{equation*}
S_k^+ = \max(0, S_{k-1}^+ + r_k - \delta)
\;\text{and}\;
S_k^- = \min(0, S_{k-1}^- + r_k + \delta)
\end{equation*}

with slack $\delta = 0.5\widehat{\sigma}$, declaring divergence when
$S_k^+ > h$ or $S_k^- < -h$ for threshold $h = 4\widehat{\sigma}$, where
$\widehat{\sigma}$ is the empirical residual scale. Accumulating the residual
rather than thresholding each sample separates persistent prediction error from
isolated deviations.

CUSUM assigns each mechanism evaluation to one of four quadrants:
\textbf{Q1} $(V_{\mathrm{obs}}\!\approx\!\widehat{V},
Y_{\mathrm{obs}}\!\approx\!\widehat{Y})$,
\textbf{Q2} $(V_{\mathrm{obs}}\!\approx\!\widehat{V},
Y_{\mathrm{obs}}\!\not\approx\!\widehat{Y})$,
\textbf{Q3} $(V_{\mathrm{obs}}\!\not\approx\!\widehat{V},
Y_{\mathrm{obs}}\!\approx\!\widehat{Y})$, and
\textbf{Q4} $(V_{\mathrm{obs}}\!\not\approx\!\widehat{V},
Y_{\mathrm{obs}}\!\not\approx\!\widehat{Y})$.
\textbf{Q1} adds supporting evidence, \textbf{Q4} contradictory evidence, and
\textbf{Q2--Q3} update edges according to their invariant causal prediction
(ICP) result~\cite{peters2016}. \S\ref{sec:koi-slow-loop} uses the \textbf{Q1} rate to adapt exploration.

\subsection{Scoring a Plan}
\label{sec:koi-cluster-objective}

Because \system scores predicted rather than known outcomes, it augments Equation~\eqref{eq:tick_program} with two terms: risk from prediction error and exploration value from testing uncertain causal relationships. Exploitation, risk, and switching cost sum over jobs; exploration is scored once per plan to avoid double-counting repeated tests. The plan score relies on a corresponding coverage assumption:

\textbf{Assumption 1.}
\label{assump:proposal-coverage}
At each tick, the planner generates $K_P$ candidate plans. Let $P_t^\star$
be the optimal feasible plan and $\Phi(\cdot)$ the plan score. We assume that,
with probability at least $\eta$, at least one candidate $P_k$ has score
\[
\Phi(P_k) \ge \Phi(P_t^\star)-\varepsilon_P,
\]
where $\varepsilon_P$ is the tolerated gap from the optimum.

\subsubsection{Exploitation}
\label{sec:exploitation}
 
Exploitation is the per-job score $J_i(L_i)$ of Equation~\eqref{eq:tchebycheff}, evaluated on the ladder's aggregated outcomes. High-confidence causal relationships guide exploitative candidates, while $J_i(L_i)$ favors those whose predicted outcomes best match the job's weighted priorities.
 
\subsubsection{Exploration}
\label{sec:exploration}
 
Exploitation alone favors deployments the surrogate already predicts to be good,
so relationships that have never been exercised stay untested. Exploration
rewards a plan for testing uncertain edges and mechanisms:
\begin{equation}
\begin{aligned}
\mathrm{Expl}(P) = \sum_{e\in E} A_e(P)\,U(\alpha_e,\beta_e)
\;+\;
\omega_m\!\!\sum_{m\in\mathcal{M}} A_m(P)\,U(\alpha_m,\beta_m),
\label{eq:koi-plan-eig}
\end{aligned}
\end{equation}
where $A_e(P)$ indicates whether any ladder in $P$ exercises edge $e$, and analogously for mechanisms. $U$ is the normalized posterior variance from before. Mechanisms, weighted by $\omega_m$, are scored separately because their scope can fail even when each constituent edge is well supported. 

\subsubsection{Risk}
\label{sec:risk}

A surrogate is most likely to be wrong when conditions have shifted, which is
also when a plan is most likely to breach an SLO. \system therefore evaluates
each prediction against a band rather than a point. Let $r_j$ be the historical
residuals for objective $j$, observed minus predicted as in
\S\ref{sec:mechanisms}, and $\widehat{\sigma}_j$ their empirical standard
deviation. The band is
\begin{equation}
\begin{aligned}
\mathrm{upper}_j
&=
\widehat{y}_j + Q_{0.95}(r_j) + 2\epsilon\widehat{\sigma}_j, \\
\mathrm{lower}_j
&=
\widehat{y}_j + Q_{0.05}(r_j) - 2\epsilon\widehat{\sigma}_j.
\end{aligned}
\label{eq:koi-dro-band}
\end{equation}
where $Q_q$ is the empirical $q$-quantile. The quantile terms carry the observed
error distribution and the $2\epsilon\widehat{\sigma}_j$ terms inflate it, with
$\epsilon$ in the role of a Wasserstein ambiguity radius around that
distribution~\cite{kuhn2025distributionally} rather than as the solution to a
distributionally robust program. \system calibrates $\epsilon$ from realized
coverage in the slow loop (\S\ref{sec:koi-slow-loop}), widening it when outcomes
fall outside the band more often than the target rate and tightening it as
calibration improves.
 
From the band, \system estimates the probability $p_j^{\mathrm{brk}}$ that each
SLO-constrained objective is violated, and combines them as
\begin{equation}
    p_i^{\mathrm{brk}}(L_i) = 1 - \prod_{j\in\mathcal{S}_i} \bigl(1 - p_j^{\mathrm{brk}}\bigr),
    \label{eq:koi-dro}
\end{equation}
where $\mathcal{S}_i$ is the set of objectives constrained by $S_i$. Treating those
objectives as independent is conservative, since latency violations are
positively correlated in practice, so the product understates the joint
satisfaction probability and overestimates risk. 
 
\subsubsection{Switching Cost}
\label{sec:switch_cost}
 
The reconfiguration cost for job $i$ is decomposed into four components:

\begin{equation*}
\mathrm{Switch}_i
=
C_{\mathrm{cold}}
+
C_{\mathrm{parallel}}
+
C_{\mathrm{kill}}
+
C_{\mathrm{risk}}.
\label{eq:koi-switch-cost}
\end{equation*}

All terms are zero for an unchanged ladder: $C_{\mathrm{cold}}$ captures loading and warm-up, $C_{\mathrm{parallel}}$ concurrent execution, $C_{\mathrm{kill}}$ draining and termination, and $C_{\mathrm{risk}}$ transient underperformance. The penalty $\lambda_t$ adapts to the observed reconfiguration rate (\S\ref{sec:koi-slow-loop}) to discourage unnecessary migrations. Separately, the swap budget $B_t$ imposes the hard \textbf{C4} cap on how many active jobs may change ladders in one tick.
 
\subsubsection{Cluster Score}
 
The four terms in Sections~\ref{sec:exploitation}--\ref{sec:switch_cost} combine into a single plan score:
\begin{equation}
\begin{aligned}
\Phi(P)
&=
\sum_{(i,L_i)\in P}
\Big[
J_i(L_i)
- \gamma\,p_i^{\mathrm{brk}}(L_i)
- \lambda_t\,\mathrm{Switch}_i(L_i^{(t-1)}, L_i)
\Big] \\
&\quad + \beta_t\,\mathrm{Expl}(P).
\end{aligned}
\label{eq:koi-cluster-score}
\end{equation}
\system aims to select the highest-scoring feasible plan,
\begin{equation}
    P_t \in \arg\max_{P\in\mathcal{F}_t} \Phi(P),
    \label{eq:koi-plan-opt}
\end{equation}
where $\mathcal{F}_t$ is the set of plans satisfying \textbf{C1--C5}. Exhaustive maximization over $\mathcal{F}_t$ is
infeasible and the per-tick search is budgeted, so \system commits the
highest-scoring feasible plan it finds rather than the maximizer
(\S\ref{sec:koi-agentic-planner}). The exploration
weight $\beta_t$ and the switching penalty $\lambda_t$ are adapted by the slow
loop (\S\ref{sec:koi-slow-loop}); the risk weight $\gamma$ is fixed.

\subsection{Self-Calibration}
\label{sec:koi-slow-loop}
 
\system adapts the quantities that govern planning from how well its predictions
have held. Edge and mechanism confidence update every tick. On a slower
schedule, \system adjusts the exploration weight $\beta_t$, the swap budget
$B_t$, the switching penalty $\lambda_t$, the robustness radius $\epsilon_t$,
the ideal point $z_t^\star$, and the CUSUM thresholds, each from accumulated
residuals.
 
Most of these follow one signal. Let $q_t$ be the fraction of decided
evidence--mechanism pairs over a rolling window of $W$ ticks that fall in
\textbf{Q1}, and with $q^\star = 1$ let the regret signal be
\begin{equation*}
    s_t = \frac{1}{W} \sum_{k=t-W+1}^{t} \max(0,q^\star-q_k),
    \label{eq:koi-regret}
\end{equation*}
which increases as predictions become less reliable. Regret is derived from the Q1 rate: a high regret signal indicates poor configurations and motivates continued exploration. The exploration weight is given by
\begin{equation*}
    \beta_{t+1} = \operatorname{clip}\bigl(
      \beta_t \exp(\eta_\beta(s_t-s^\star)),\;
      \beta_{\min},\; \beta_{\max}
    \bigr),
    \label{eq:koi-beta-update}
\end{equation*}
with target regret $s^\star$ and adaptation rate $\eta_\beta$. The same signal
raises the swap budget $B_t$ when predictions are unreliable, so more active jobs
may be reconsidered. The switching penalty $\lambda_t$ instead follows the
observed reconfiguration rate, rising as churn grows to damp thrashing.
 
The radius $\epsilon_t$ follows realized coverage, widening when outcomes fall
outside the band more often than the target rate and tightening otherwise. The
ideal reference tracks the best outcomes a job has achieved,
\begin{equation*}
z_{i,j,t}^\star=
\begin{cases}
\displaystyle
\max_{r\in\mathcal{H}_{i,t}} y_{r,j}+\Delta_j, & \text{$j$ maximized},\\[8pt]
\displaystyle
\min_{r\in\mathcal{H}_{i,t}} y_{r,j}-\Delta_j, & \text{$j$ minimized},
\end{cases}
\label{eq:koi-zstar-update}
\end{equation*}
where $\mathcal{H}_{i,t}$ is job $i$'s deployment history and $\Delta_j$ a
margin that preserves dominance. The self-calibration loop relies on a corresponding convergence assumption:

\textbf{Assumption 2 (Convergence).}
\label{assump:convergence}
If jobs, workloads, available capacity, and serving conditions remain stable
for sufficiently long, we assume that \system converges to a stable operating
point. At this point, predictions remain accurate, exploration no longer finds
materially better plans, reconfiguration diminishes, and the learned causal
and robustness state changes only marginally.

\subsection{Agentic Planner}
\label{sec:koi-agentic-planner}

 The search space for a cluster deployment is vast and coupled. Each chain sets about 140 knobs (Table~\ref{tab:knobs}), giving at least $2^{140}$ configurations per chain before replica counts and multi-rank ladders are considered. A plan chooses one ladder for every job, so the number of candidate plans grows exponentially with the number of jobs, and the choices are coupled because jobs with different workloads and SLOs compete for the same capacity. \system therefore never enumerates this space. Budget books and backward DAG queries, described next, narrow it for each job, an LLM proposes a small set of ladders within it, and deterministic code evaluates, scores, and selects among them.
 
A root planner receives the frozen cluster snapshot and assigns each job a \emph{budget book}, a soft upper bound on which GPU types and instance pools the job may consider. Budgets can overlap in resources selected, and contention is resolved later by a capacity-constrained solver that assigns at most one ladder per job to the available GPUs. The root planner invokes per-job specialist sub-agents to propose candidate ladders within each job's budget. Each specialist sub-agent receives the job's model architecture, workload profile, current deployment, budgeted hardware, and all mechanisms whose scope matches the job. It can query the DAG backward: given a target SLO, the sub-agent traces which $X$ decisions affect it through which $V$ mediators, focusing on the decisions that matter (Section~\ref{sec:mechanisms}). This division lets the root reason about cluster-wide tradeoffs (which scarce accelerators to reserve, which workloads can share resources) while sub-agents focus on the configuration details of individual models.

These proposals are advisory: \system first validates each candidate for physical feasibility, then invokes the surrogate to predict and score its performance. The resulting ladders compete for finite capacity under the joint selector, which assigns at most one ladder per job to maximize cluster-wide gain (Section~\ref{sec:koi-cluster-objective}) while accounting for unserved demand, improvement over active deployments, and switching cost. The root planner may inspect the result and propose additional candidates, but cannot override deterministic scores, capacity checks, or constraints \textbf{C1--C5}.

\subsection{Implementation}

Batch and online inference jobs are executed through two different pathways. For batch jobs, large prompt files are sharded and a custom shard manager coordinates shard assignment and completion for each worker. This ensures that every rank and chain planned by \system make progress on the same job without conflicts. Online jobs are deployed through the Nvidia Dynamo inference deployment framework. \system decides on the exact chain configuration and desired number of maximum replicas, while Dynamo autoscales within that maximum replica bound in response to request load and desired SLAs. Per-rank telemetry from all running jobs is continuously streamed from the different worker clusters to a unified cluster state store. At each five-minute tick, \system reads a snapshot of this store to use for the next plan.

\begin{table*}[tbp]
  \centering
  \caption{Simulated benchmark results for 3-hour (400-GPU) and 12-hour (800-GPU) heterogeneous workloads. \textbf{Bold} marks the best value and \underline{underline} the second-best; ties are marked for all matching methods}
  \label{tab:simulation-results}
  \scriptsize
  \setlength{\tabcolsep}{3pt}
  \begin{tabularx}{\textwidth}{@{}p{0.145\textwidth}@{\hspace{4pt}}c*{7}{Y}|Y@{}}
    \toprule
    Method & Hrs &
    \shortstack{Batch\\prog.\ (\%)} &
    \shortstack{Admit\\online (\%)} &
    \shortstack{Admit\\batch (\%)} &
    \shortstack{\$/token\\($10^{-4}$)} &
    \shortstack{TTFT\\met (\%)} &
    \shortstack{TPOT\\met (\%)} &
    \shortstack{Goodput\\(tok/s)} &
    \shortstack{Overall\\score} \\
    \midrule
    \multicolumn{10}{@{}l}{\textit{Non-learning baselines}}\\
    \midrule
    \multirow{2}{=}{FCFS + first-fit}
      & 3  & \underline{57.71} & \underline{77.78} & \underline{57.14} & 2.95 & 3.90 & 6.89 & 1830.84 & 49.13 \\
      & 12 & 38.69 & \underline{78.26} & 50.00 & 6.09 & 2.45 & 6.27 & 1772.01 & 31.46 \\
    \addlinespace[2pt]
    \multirow{2}{=}{Best fixed configuration}
      & 3  & \underline{57.71} & 66.67 & \underline{57.14} & 2.91 & 5.24 & 13.61 & 1857.31 & 50.24 \\
      & 12 & 24.54 & 69.57 & 57.14 & 29.65 & 7.63 & 13.55 & 364.03 & 17.59 \\
    \addlinespace[2pt]
    \multirow{2}{=}{EDF + first-fit}
      & 3  & \underline{57.71} & \underline{77.78} & \underline{57.14} & 2.94 & 4.05 & 7.27 & 1833.16 & 49.27 \\
      & 12 & 38.62 & \underline{78.26} & 50.00 & 6.09 & 2.36 & 6.21 & 1771.49 & 31.40 \\
    \addlinespace[2pt]
    \midrule
    \multicolumn{10}{@{}l}{\textit{Placement and search baselines}}\\
    \midrule
    \multirow{2}{=}{Rolling CP-SAT}
      & 3  & 41.99 & 72.22 & \underline{57.14} & 6.69 & 3.00 & 15.88 & 807.03 & 31.90 \\
      & 12 & 14.99 & \underline{78.26} & 64.29 & 40.18 & 2.89 & 16.83 & 268.57 & 15.83 \\
    \addlinespace[2pt]
    \multirow{2}{=}{Contextual bandit}
      & 3  & 44.42 & 66.67 & \underline{57.14} & 5.43 & 8.09 & \underline{27.08} & 993.08 & 37.49 \\
      & 12 & 30.52 & 65.22 & 50.00 & 14.16 & 9.56 & 23.06 & 762.20 & 24.89 \\
    \addlinespace[2pt]
    \multirow{2}{=}{PRISM}
      & 3  & 42.55 & \underline{77.78} & \underline{57.14} & 7.06 & 2.80 & 5.16 & 764.66 & 29.06 \\
      & 12 & 15.35 & 73.91 & 64.29 & 57.66 & 1.94 & 4.69 & 187.18 & 11.14 \\
    \addlinespace[2pt]
    \multirow{2}{=}{Rolling EvoX}
      & 3  & \underline{57.71} & 66.67 & \underline{57.14} & \underline{2.69} & \underline{9.90} & 22.10 & \underline{2004.06} & \underline{54.34} \\
      & 12 & \textbf{50.71} & 65.22 & 64.29 & \textbf{3.37} & \underline{11.96} & \underline{24.02} & \underline{3012.00} & \underline{50.01} \\
    \addlinespace[2pt]
    \multirow{2}{=}{OpenEvolve}
      & 3  & 56.70 & 66.67 & \underline{57.14} & 4.74 & 5.22 & 23.41 & 1139.00 & 41.74 \\
      & 12 & 25.01 & \underline{78.26} & 64.29 & 26.24 & 9.19 & 16.14 & 411.25 & 19.84 \\
    \addlinespace[2pt]
    \multirow{2}{=}{GPT-Only (non-agentic)}
      & 3  & 52.04 & \underline{77.78} & \textbf{71.43} & 6.31 & 3.98 & 18.98 & 855.60 & 35.28 \\
      & 12 & 17.54 & 69.57 & \underline{71.43} & 16.86 & 3.02 & 17.29 & 640.25 & 21.34 \\
    \addlinespace[2pt]
    \multirow{2}{=}{\system}
      & 3  & \textbf{84.55} & \textbf{94.44} & \textbf{71.43} & \textbf{1.85} & \textbf{28.79} & \textbf{41.80} & \textbf{2910.64} & \textbf{70.83} \\
      & 12 & \underline{49.61} & \textbf{95.65} & \textbf{100.00} & \underline{3.39} & \textbf{27.95} & \textbf{35.80} & \textbf{3184.15} & \textbf{63.13} \\
    \midrule
    \multirow{2}{=}{\textbf{$\Delta$ vs.\ best baseline}}
      & 3  & +26.84pp & +16.66pp & +0.00pp & +31.23\% & +18.89pp & +14.72pp & +45.24\% & +30.35\% \\
      & 12 & $-1.10$pp & +17.39pp & +28.57pp & $-0.59$\% & +15.99pp & +11.78pp & +5.72\% & +26.23\% \\
    \bottomrule
  \end{tabularx}
\end{table*}

\begin{table*}[tbp]
  \centering
  \caption{Real-hardware results for the 3-hour, 104-GPU heterogeneous benchmark. \textbf{Bold} marks the best value and \underline{underline} the second-best.}
  \label{tab:real_cluster-results}
  \scriptsize
  \setlength{\tabcolsep}{4pt}
  \begin{tabularx}{\textwidth}{@{}p{0.17\textwidth}@{\hspace{5pt}}*{7}{Y}|Y@{}}
    \toprule
    Method &
    \shortstack{Batch\\prog.\ (\%)} &
    \shortstack{Admit\\online (\%)} &
    \shortstack{Admit\\batch (\%)} &
    \shortstack{\$/token\\($10^{-4}$)} &
    \shortstack{TTFT\\met (\%)} &
    \shortstack{TPOT\\met (\%)} &
    \shortstack{Goodput\\(tok/s)} &
    \shortstack{Overall\\score} \\
    \midrule
    Contextual bandit & 50.00 & 53.33 & \textbf{100.00} & 0.45 & 22.23 & 33.66 & 2291.03 & 51.05 \\
    Rolling EvoX      & 50.00 & \underline{80.00} & \textbf{100.00} & 0.44 & 28.08 & 35.48 & 2348.15 & 54.90 \\
    OpenEvolve        & \underline{75.00} & \textbf{100.00} & \textbf{100.00} & \underline{0.30} & \underline{65.66} & \underline{91.41} & \underline{3497.43} & \underline{84.44} \\
    \system           & \textbf{100.00} & \textbf{100.00} & \textbf{100.00} & \textbf{0.24} & \textbf{79.05} & \textbf{99.11} & \textbf{4119.51} & \textbf{94.50} \\
    \midrule
    \textbf{$\Delta$ vs.\ best baseline} & +25.00pp & +0.00pp & +0.00pp & +20.00\% & +13.39pp & +7.70pp & +17.79\% & +11.91\% \\
    \bottomrule
  \end{tabularx}
\end{table*}

\section{Evaluation}
\label{sec:evaluation}
Our evaluation asks three questions. First, how does \system compare with static, classical, outcome-learning, evolutionary, and language-model planners in goodput, SLO attainment, admission, and cost (\S\ref{sec:evaluation-end-to-end})? Second, do joint optimization, evidence accumulation, and health-triggered recovery operate as intended, and how do these effects change with deployment scale and duration (\S\ref{sec:evaluation-design-action})? Third, what do deterministic planning tools, the causal evidence graph, and cross-tick learning contribute, and how sensitive are the results to the language-model backbone (\S\ref{sec:evaluation-ablation})?

\subsection{Experimental Setup}

\textbf{Benchmarks.}
We evaluate \system in one real-hardware setting and three simulated settings: a 3-hour deployment on a 104-GPU heterogeneous GCP cluster, 3-hour and 12-hour heterogeneous simulations, and a 12-hour homogeneous-cluster simulation. Each setting defines a benchmark consisting of a shared resource pool and a concurrent mix of online and batch workloads. For each benchmark, \system and each of the ten baselines independently generate the placements and configurations used to serve the same workload.

Each benchmark mixes 25 online and batch workloads across ten model architectures: Llama-3.2-1B, Llama-3.3-70B, Qwen2.5-Coder-32B, Phi-4, Mixtral-8x7B, Kimi-K2-Instruct, GLM-4.5-Air, Qwen3-Next-80B, DeepSeek-V3 and Qwen3-235B-A22B-Instruct, spanning 1B to 235B parameters and including dense, mixture-of-experts, and frontier models. Online workloads distribution range from short moderation (256/8 tokens, TTFT $<100$ ms) to long-context summaries (24,000/1,200 tokens, TTFT $<3$s) and decode-heavy agentic tasks (1,400/4,096 tokens). Six batch jobs carry fixed deadlines ranging from 20 to 135 minutes. Request traffic follows Poisson, constant, and gamma arrival processes parameterized from production profiles to each job type. The workloads used in simulation and real cluster are summarized in  Figure~\ref{fig:benchmark-timeline} and Table~\ref{fig:benchmark_workloads} in the Appendix.

\textbf{Real-Hardware Execution.}
Each experiment is conducted with three Google Cloud Kubernetes clusters across three distinct regions: us-central1, us-east5 and europe-west2. There are 104 GPUs in total, consisting of a heterogeneous mix of H100-80GB, A100-80GB, A100-40GB, and RTXPRO6000. Due to cost constraints, non-evolutionary algorithm based baselines were omitted from real experiments. 

\textbf{Simulation Setup.}
For the simulation environment, we used LLMServingSim~2.0~\cite{11527300}, a discrete event simulator (DES) as the execution environment for the simulated clusters. Unlike Koi's internal surrogate model, LLMServingSim 2.0 does not provide point estimates per placement, but rather simulates request execution and queueing semantics for all requests routed to a deployed configuration. After a plan is selected by Koi, LLMServingSim 2.0 executes the corresponding configuration and provides the realized performance observations that would otherwise come from hardware telemetry. It simulates using per-operator compute profiles across GPU, model, and tensor-parallel configurations, and an ASTRA-SIM \cite{rashidi2020astrasim} backend for collective communication. To support all model–GPU combinations in our benchmark, we augmented LLMServingSim 2.0 with layer-wise profiles for the additional LLM models and GPU types used in the evaluation. 

\textbf{Baselines.}
\system is compared against ten schedulers, grouped by the capability each baseline is intended to test.

\emph{Non-learning baselines.} FCFS+First-Fit admits jobs in arrival order and
assigns the least-VRAM feasible placement. EDF~\cite{liu1973scheduling} applies
the same placement rule but prioritizes earlier deadlines. Best Fixed uses one
static hardware and parallelism configuration per job, selected offline by
GPT-5.6-Sol with access to the complete workload trace; at runtime, it reuses
these configurations and falls back to VRAM-based placement when the selected
configuration is unavailable.

\emph{Per-job optimizers.} 
Prism~\cite{prism} estimates memory and KV-cache pressure to place preconfigured model-parallel groups, reallocating memory at runtime but never choosing GPU type, parallelism, or replica count. To adapt Prism to cluster scheduling, we enumerate policy-compliant GPU-pool and TP/PP placement candidates using workload-derived VRAM and KV-cache demand. Prism then admits waiting jobs by assigning high-KV-demand workloads to the least-pressured feasible pool under global capacity constraints.

\emph{Global optimizers without mechanism learning.} Rolling CP-SAT~\cite{perron2023cpsatlp} solves ILP at each tick over the same candidate catalog and constraints as \system. It enumerates policy-valid placements using checked-in maximum VRAM requirements and resource-map capacity, then optimizes lexicographically for priority-weighted admission, lower estimated reservation cost, and lower GPU footprint. But it never learns from outcomes. 
Contextual Thompson sampling~\cite{agrawal2013thompson} maintains a Bayesian linear reward model over public job and candidate-placement features, updating only from terminal deployment outcomes. Its bounded reward combines online TTFT, TPOT, and throughput attainment with weights $0.4$, $0.4$, and $0.2$, respectively; for batch jobs, it combines throughput attainment and deadline completion with weights $0.4$ and $0.6$.

\emph{LLM-guided planners.} GPT-only (non-agentic) receives a frozen cluster snapshot, containing the same public job, resource, policy, and model-catalog information available to \system, and generates a placement plan without tools or cross-tick state. Its prompt requires a complete policy-compliant action for each job, specifying feasible GPU pool and TP/PP/DP configuration while respecting current capacity. Agentic additionally uses a REPL and the same read-only inspection and placement-validation tools available to \system's agent.

\emph{Evolutionary search.} Rolling EvoX~\cite{liu2026evox} evolves a deterministic placement function at each tick, adapting its search strategy when progress stalls. OpenEvolve~\cite{sharma2025openevolve} retains a MAP-Elites database of diverse, high-scoring placement programs across ticks. In each baseline, we used the similar objective function as \system to make it comparable. EvoX's each reasoning round takes $\approx$ 30 seconds. Considering reasonable scheduling overhead, we set it to max 8 rounds or 5 minutes whatever returns first. Again, we provided the same information with both EvoX and OpenEvolve baselines that was given to \system.


For the main performance evaluation, we used the same LLM model (GPT-5.6-Luna) for the policies that uses LLM. Additionally, we did LLM-backbone sensitivity study with different LLM models for \system. To standardize the comparison, each baseline is given the same cluster snapshots, candidate catalog, and safety checks. Differences are limited to the plan-search policy and, for \system alone, cross-tick mechanism learning.

\textbf{Metrics.}
A single evaluator computes every number from per-run traces. All denominators belong to the workload (offered requests, submitted jobs, full campaign sizes, a common clock $T$), so no planner can improve a metric by refusing, dropping, or narrowing work. Stochastic planners report the mean and range over at least three runs.

We collapse results into an overall score with three equally weighted pillars: Quality (TTFT SLO met, TPOT SLO met, deadline met), Coverage (online admission, batch admission, batch progress), and Efficiency (goodput and cost per SLO-token, each normalized to the best in each comparison). The score is their geometric mean; no pillar compensates for another. The problem formulation uses the same instantaneous outcome vector for online and batch jobs (Section~\ref{sec:problem-formulation}). The evaluation, however, judges batch jobs by completion progress and deadline attainment, which capture finite-work completion rather than instantaneous serving quality.

\emph{SLO met} is the fraction of offered online requests meeting both TTFT and TPOT p99 targets, averaged per job so one high-volume tenant cannot dominate. \emph{Deadline met} is the share of batch jobs finishing before their deadline. \emph{Admission} is the share of submitted jobs whose ranks materialized without a fatal interruption. \emph{Batch progress} is request-weighted completion over all submitted jobs. \emph{Goodput} is SLO-honoring output tokens per second: online tokens passing both latency gates plus the projected tokens of deadline-meeting jobs. \emph{Cost per SLO-token} divides the fleet capacity cost by total SLO-token output; idle GPUs count as waste, so the cheapest planner produces the most SLO output from the same sunk fleet. Planner LLM cost is reported separately.

\subsection{End-to-End Performance}
\label{sec:evaluation-end-to-end}

\paragraph{End-to-End Performance overview.}
Table~\ref{tab:simulation-results} reports the end-to-end results. Over three hours, \system achieves 2,911 goodput tok/s, 45.24\% more than the strongest baseline, while reducing cost per token by 31.23\%; it also improves batch progress by 26.84\% and TTFT/TPOT SLO attainment by 18.89/14.72\%. Over twelve hours, \system admits 95.65\% of online jobs and all batch jobs, improves goodput by 5.72\%, and leads TTFT/TPOT attainment by 15.99/11.78\%, while batch progress is 1.10 points below EvoX and cost is 0.59\% higher. Thus, \system's gain does not come from maximizing one workload class: it sustains more admitted work and substantially better online latency while remaining competitive on batch progress and cost.

\subsection{Design Evaluation}
\label{sec:evaluation-design-action}

\paragraph{Joint optimization.} In the real-cluster run, a Qwen3-235B job waits at tick 18 while ten services are running. The only reclaim candidate would reduce a Phi-4 service from two TP2 RTXPRO6000 replicas to one. This would halve its modeled capacity from 1,572 to 786 tok/s, below its 834 tok/s requirement, while its observed p99 TTFT already exceeds the 500\,ms target at 645\,ms. A full preemption would be worse, reducing that service's availability. The joint selector rejects the reclaim as not Pareto-better than keeping Phi-4, defers Qwen for two ticks, and admits it at tick 20 after another batch finishes naturally. The ten-minute delay protects an existing service from a predicted SLO regression without sacrificing Qwen's eventual admission.

\paragraph{Deployment evidence makes planning more calibrated.}
In the 12-hour simulation on heterogeneous cluster, the mean causal uncertainty $\bar U$ falls from a first-quarter median of 0.0215 to 0.00053 in the final quarter, a $41\times$ reduction. The fraction of predictions receiving a learned residual correction rises from 2.3\% to 10.8\%, while their median absolute log error falls from 0.965 to 0.823. During warm-up, the regret signal raises the exploration weight from 0.49 to 2.0 by tick 23. As evidence accumulates, the median DRO radius tightens from 0.379 to 0.260, then widens again after the late workload shift since the uncertainty increases again with it. Thus, \system becomes more confident where observations repeat while retaining the ability to increase exploration and robustness when conditions change.

\paragraph{Swap.}
In the real cluster experiment, a Mixtral service reaches critical health at tick 15 after only one of its two A100-80GB replicas materializes and observed p99 TTFT reaches 2,299\,ms against a 700\,ms target. The per-job sub-agent proposes one physically launchable TP8 A100 rank, and the deterministic scorer assigns a $25.44$ gain over keeping the failed deployment, including a $25.00$ critical-rehabilitation credit. The joint selector commits only this swap and keeps the other active jobs unchanged. The request-level TTFT pass rate rises from 2.4\% before recovery to 99.9\% on the relaunched rank, and the job is healthy by tick 18. This example connects telemetry, candidate generation, deterministic scoring, and cluster-wide commitment in one decision.

\paragraph{Learning and performance improvement.}
We distinguish fleet scale from elapsed learning time using two comparisons. Over roughly three hours, the 104-GPU real cluster commits 20 deployments and selects eight mechanisms, whereas the 800-GPU simulation commits 80 deployments and selects 17 mechanisms. At a larger cluster scale (800 GPUs for 12 hours vs. 104 GPUs for 3 hours), \system gathered more evidence, increasing from 407 to 2,272 pieces of evidence, and ran more CUSUM tests, increasing from 18,705 to 101,420. The share of decisive mechanism tests that support their predicted mediator--outcome path rises from 1.9\% to 19.1\%, and ICP acceptance for the same workload-to-mediator edge rises from 11.0\% to 63.3\%. These runs show that more concurrent deployments activate learning sooner and that a longer horizon strengthens its evidence. 

\begin{figure}[t]
  \centering
  \includegraphics[
    width=\columnwidth,
    keepaspectratio
  ]{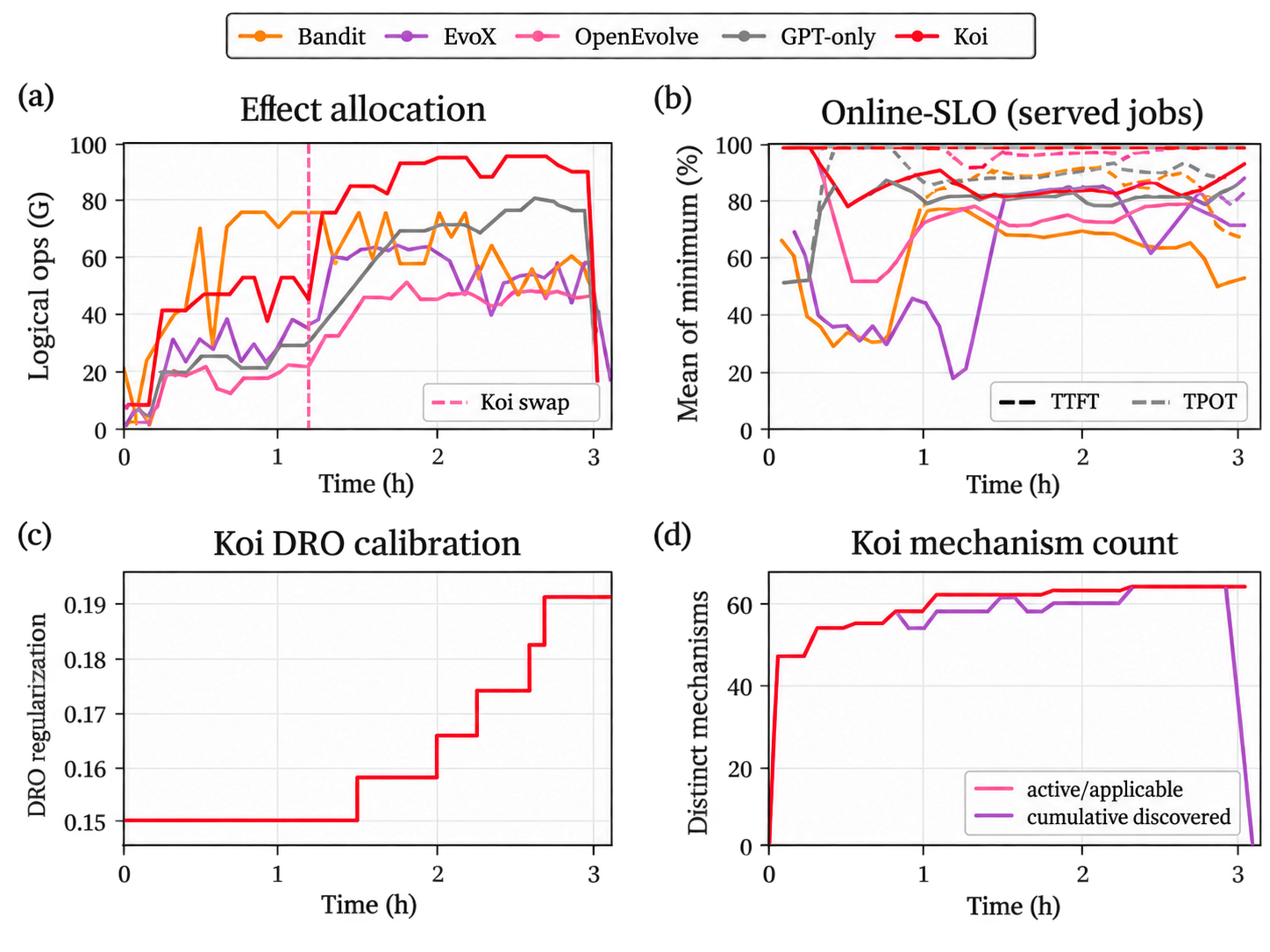}
    \caption{\textbf{Real-cluster allocation, SLO attainment, and adaptation over three hours.} (a) Plan-declared logical GPU allocation for each policy, with \system's realized swap marked; (b) equal-job mean TTFT (solid) and TPOT (dashed) attainment among served online jobs; (c) \system's calibrated DRO radius; and (d) active/applicable and cumulatively discovered mechanisms.}
  \label{fig:reconfiguration}
  \Description{\gmnew{Four time-series panels from the three-hour real-cluster runs compare logical GPU allocation and served-job SLO attainment across five policies, and show \system's DRO radius and active and cumulative mechanism counts.}}
\end{figure}

\paragraph{Evidence changes both allocation and planner state.}
Figure~\ref{fig:reconfiguration} shows \system increasing its plan-declared allocation above 90\% as jobs arrive while sustaining high served-job TTFT and TPOT attainment. As observed outcomes expose prediction-band misses, \system widens its DRO radius from 0.15 to 0.191 rather than continuing to trust an under-covering bound. In parallel, the cumulative mechanism set grows to 64 as new workload contexts arrive.

\begin{figure}[t]
\centering
\includegraphics[width=\columnwidth]{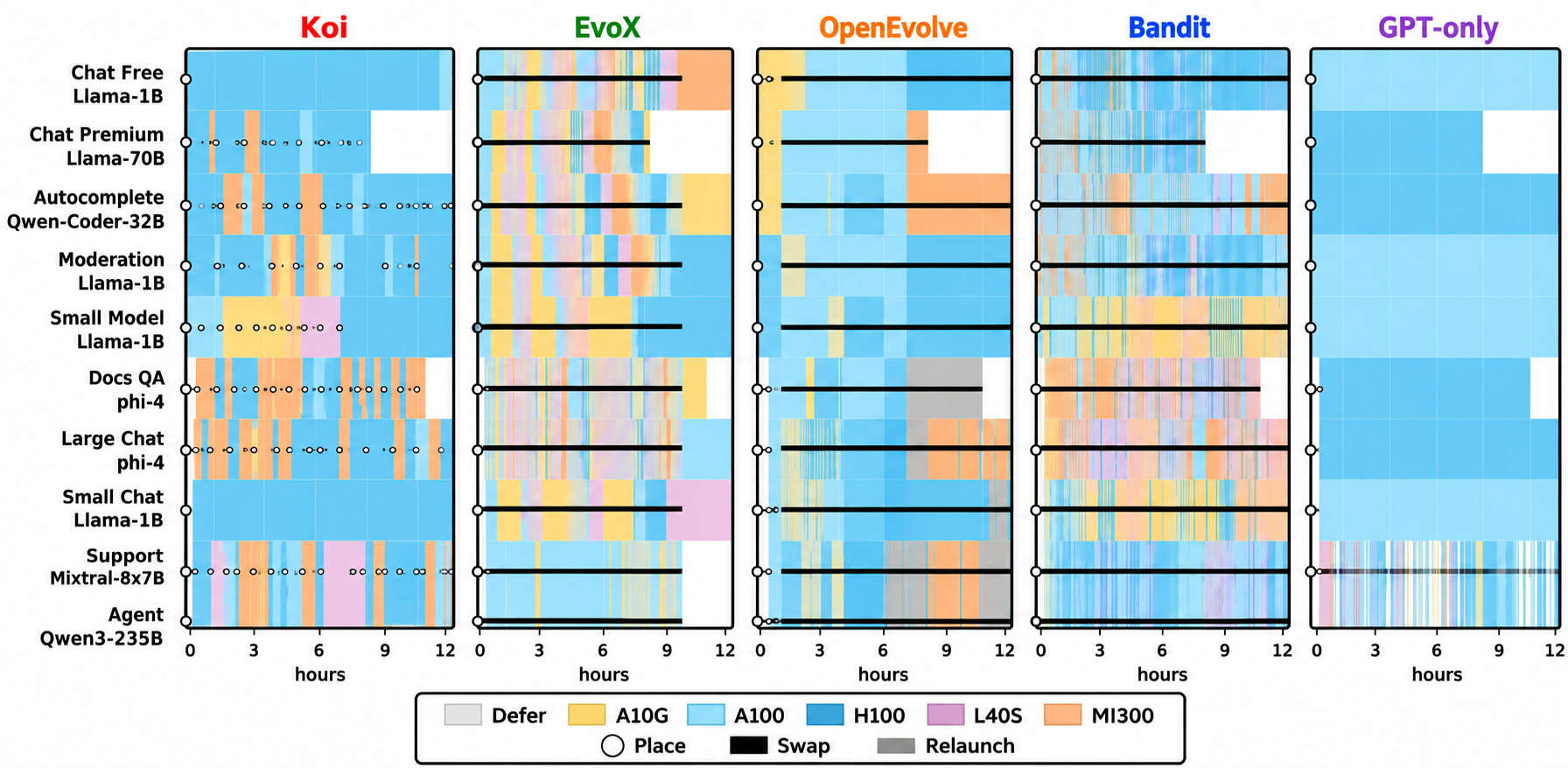}
\caption{\textbf{Online workload scheduling across methods.}
Per-job GPU assignments for nine workloads over 12 hours under
\system, EvoX, OpenEvolve, the contextual bandit, and GPT-only.
Colors denote GPU families; markers indicate placement, swap,
and rank relaunch events.}
\label{fig:selected-online-timeline}
\Description{Five aligned timelines compare GPU assignments for
the same nine online workloads over 12 hours. Rows identify
workloads, columns represent time, colors identify GPU families,
and markers identify placement, swap, and rank relaunch events.}
\end{figure}

\paragraph{Balance between exploitation and exploration.}
Figure~\ref{fig:selected-online-timeline} shows long stable placement intervals under \system, with changes concentrated on a subset of degraded services. EvoX and the contextual bandit instead produce dense swap sequences across nearly every online workload, while OpenEvolve and GPT-only perform no explicit swaps and largely preserve their initial choices; OpenEvolve's additional transitions are failure-driven relaunches. Across the complete trace, \system issues 201 swaps and 2,076 keeps, compared with 1,464 swaps for EvoX and 1,928 for the bandit. The timeline therefore places \system between two failure modes: repeatedly perturbing productive deployments and never correcting a poor initial placement.

\begin{figure}[t]
  \centering
  \includegraphics[
    width=\columnwidth,
    keepaspectratio
  ]{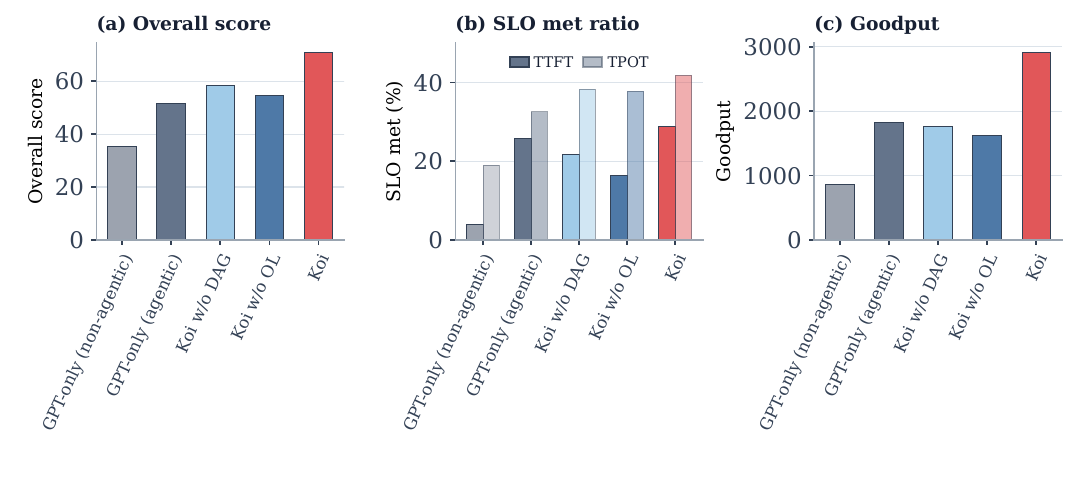}
  \caption{\textbf{Ablations over the 3-hour heterogeneous simulation.} Overall score, TTFT and TPOT SLO attainment, and goodput across five planner variants.}
  \label{fig:mechanism-ablation}
  \Description{Three bar-chart panels compare five planner variants on overall score, TTFT and TPOT SLO attainment, and goodput. Full \system achieves the highest value across all reported metrics, while GPT-only and \system ablations perform worse.}
\end{figure}

\subsection{Ablation Study}
\label{sec:evaluation-ablation}

\paragraph{\system sub-component ablation.}
Figure~\ref{fig:mechanism-ablation} and Table~\ref{tab:appendix-ablation-results} compare progressively more complete planners. Giving the language model deterministic prediction, scoring, and validation tools increases goodput from 856 to 1,827 tok/s, raises TTFT attainment from 3.98\% to 25.74\%, and reduces cost per token from $6.31\times10^{-4}$ to $2.95\times10^{-4}$. The complete system reaches 2,911 tok/s; removing the causal graph reduces goodput to 1,770 tok/s, while freezing online updates reduces it to 1,621 tok/s. Full \system also completes 84.55\% of batch work, versus 59.42\% without the graph and 63.01\% without online learning, despite admitting fewer batch jobs than either variant. The result separates two roles: deterministic tools ground each planning attempt, while persistent mechanism evidence improves which valid plans are retained and revisited across ticks.

\paragraph{LLM backbone ablation.} 
The backbone ablation localizes the Cohere--GPT-5.6 Sol difference to a single
7,200-request Llama-70B batch campaign. Both planners admit it at 00:15 and never reconfigure it, but Cohere selects one eight-GPU MI300 TP8 chain and finishes at 01:07 with 82 minutes of deadline slack, whereas Sol selects four TP4 A100 replicas using 16 GPUs and reaches 7,115 requests by the 02:30 deadline, missing by 85. Because batch goodput credits the campaign only when it meets its deadline, this single decision contributes 8.64 million tokens and explains 99.3\% of their trace-derived goodput gap; their online contributions differ by only 57,921 tokens. The insight is that deterministic validation bounds the action space but cannot make all valid candidates equally useful: a small number of initial proposals can dominate the final result. This single-run comparison does not isolate accelerator type, parallelism, replication, or batching limits, which change together.
Another insight is that the achieved outcomes (TTFT, TPOT etc.) of all the jobs, in all ablations are well within three standard deviations of each other - indicating a model agnostic behavior.

\section{Discussion, Limitations and future work}
\label{sec:discussion}

\system benefits from a larger fleet as it
runs more concurrent jobs, exercising more relationships per tick and covering
more scopes. Evidence accumulates faster and applies to a wider set of future
jobs. Additional headroom also makes exploration affordable without putting an
SLO at risk, and planner cost is charged per tick and amortized over the jobs in
it. Utilization rises 24.5\% between the first and last quarter of the 12-hour
run, but the run is too short to reach the convergence conditions. New decision axes enlarge $X$ without changing \system's per-tick cost, which is
bounded by the number of candidates evaluated rather than by the size of the
space. Expert parallelism, prefill--decode disaggregation, and engine-level
tuning are therefore straightforward to admit. Each requires only that telemetry
report at least one mediator through which its effect passes, since an axis with
no measured mediator can be scored but not tested. Enumerative and black-box
search have no such property and require more samples as the space grows. New
accelerators, including AMD and TPU can then be additional environments in
$R_t$ and need no change to the planner. Training, post-training, and fine-tuning jobs have the same three properties
\system relies on: a configuration space that can be enumerated, runtime
behavior the stack already reports, and evaluations too expensive to run
exhaustively. Extending the planner to admit them would let one controller
allocate a fleet across all of these workloads rather than inference alone.

\section{Conclusion}

Heterogeneous LLM inference requires more than a better predictor.  A useful
controller must allocate scarce devices globally, act safely under prediction
error, and decide which expensive deployments should improve its future
knowledge.  \system addresses these requirements by attaching falsifiable,
mediator-grounded mechanisms to deployment plans; using a
sandboxed LLM only to propose candidate plans; and keeping scoring, validation,
and side effects deterministic.  Telemetry closes a calibrated slow loop, while
a stateless execution plane keeps fast serving and recovery independent of the
planner.  The result is a cluster control architecture that can improve with
use without making availability depend on an agent's correctness.

\section{Acknowledgements}
Generative AI tools, specifically ChatGPT, Claude Code and DeepSeek, were used to assist with generating sections of this work, including text, tables, graphs and code.


\bibliographystyle{ACM-Reference-Format}
\bibliography{ms.bib}

\appendix

\begin{table*}[tbp]
  \centering
  \caption{Mixed online/batch homogeneous results. All runs use the
  12-hour workload horizon on the simulator.}
  \label{tab:appendix-mixed-results}
  \scriptsize
  \setlength{\tabcolsep}{4pt}
  \begin{tabularx}{\textwidth}{@{}p{0.20\textwidth}*{5}{Y}|Y@{}}
    \toprule
    Method &
    \shortstack{Batch\\progress (\%)} &
    \shortstack{Admit (\%)\\online / batch} &
    \shortstack{\$/token\\($10^{-4}$)} &
    \shortstack{SLO met (\%)\\TTFT/TPOT} &
    \shortstack{Goodput\\(tok/s)} &
    \shortstack{Overall\\score} \\
    \midrule
    Best Fixed Configuration & 37.58 & 60.87/42.86 & 3.56 & 24.38/31.09 & 2147.77 & 50.69 \\
    Contextual Bandit & 1.89 & 43.48/7.14 & 14.04 & 10.32/10.95 & 544.34 & 14.59 \\
    Rolling EvoX & 14.92 & 65.22/14.29 & 4.76 & 29.61/37.10 & 1606.39 & 40.37 \\
    Openevolve & 14.92 & 65.22/14.29 & 6.76 & 20.26/32.46 & 1130.77 & 33.18 \\
    GPT-Only (non agentic) & \textbf{40.06} & 65.22/\textbf{57.14} & 4.29 & 18.15/33.62 & 1783.61 & 49.17 \\
    \system & 36.62 & \textbf{78.26}/\textbf{57.14} & \textbf{3.51} & \textbf{31.37}/\textbf{39.29} & \textbf{2175.63} & \textbf{56.68} \\
    \midrule
   \textbf{$\Delta$ vs.\ best baseline} & $-3.44$pp & +8.11pp & +1.40\% & +1.76pp / +2.19pp & +1.30\% & +11.82\% \\
    \bottomrule
  \end{tabularx}
\end{table*}

\begin{table*}[tbp]
  \centering
  \caption{Agentic and \system\ ablations. All runs use the 3-hour workload horizon on the simulator.}
  \label{tab:appendix-ablation-results}
  \scriptsize
  \setlength{\tabcolsep}{4pt}
  \renewcommand{\arraystretch}{1.2}
  \begin{tabularx}{\textwidth}{@{}p{0.28\textwidth}*{5}{Y}|Y@{}}
    \toprule
    Variant &
    \shortstack{Batch\\progress (\%)} &
    \shortstack{Admit (\%)\\online / batch} &
    \shortstack{\$/token\\($10^{-4}$)} &
    \shortstack{SLO met (\%)\\TTFT/TPOT} &
    \shortstack{Goodput\\(tok/s)} &
    \shortstack{Overall\\score} \\
    \midrule
    GPT-only (non agentic)  & 52.04 & 77.78/71.43 & 6.31 & 3.98/18.98 & 855.60 & 35.28 \\
    Opencode SDK (GPT-5.6-Luna)  & 54.77 & 88.89/57.14 & 2.95 & 25.74/32.62 & 1826.64 & 51.55 \\
    \system without DAG/mechanisms & 59.42 & 94.44/100.00 & 3.05 & 21.66/38.33 & 1770.03 & 58.34 \\
    \system without online learning & 63.01 & 94.44/85.71 & 3.33 & 16.49/37.74 & 1621.48 & 54.79 \\
    Full \system & \textbf{84.55} & \textbf{94.44}/71.43 & \textbf{1.85} & \textbf{28.79}/\textbf{41.80} & \textbf{2910.64} & \textbf{70.83} \\
    \midrule
    \textbf{$\Delta$ vs.\ best baseline} & +21.54pp & $-8.00$pp & +37.29\% & +3.05pp & +59.34\% & +21.41\% \\
    \bottomrule
  \end{tabularx}
\end{table*}

\begin{table}[tbp]
  \centering
  \caption{\system\ backbone LLM ablation. All runs use the 3-hour workload horizon on the simulator.}
  \label{tab:appendix-backbone-ablation}
  \scriptsize
  \setlength{\tabcolsep}{3pt}
  \begin{tabularx}{\columnwidth}{@{}p{0.26\columnwidth}*{3}{Y}|Y@{}}
    \toprule
    Base model &
    \shortstack{Batch\\progress (\%)} &
    \shortstack{Admit (\%)\\online / batch} &
    \shortstack{Planner\\\$/tick} &
    \shortstack{Overall\\score} \\
    \midrule
    GPT-5.6 Sol         & 67.31 & 94.44/85.71 & 0.75 & 62.12 \\
    DeepSeek-V4 Pro     & 48.76 & 94.44/71.43 & 0.30 & 50.44 \\
    Cohere Command-A    & 67.37 & 94.44/85.71 & 0.44 & 68.84 \\
    \midrule
    \system\ (default, GPT-5.6 Luna) & \textbf{84.55} & \textbf{94.44}/71.43 & \textbf{0.057} & \textbf{70.83} \\
    \bottomrule
  \end{tabularx}
\end{table}

\begin{figure*}[!t]
\centering

\begin{minipage}[t]{0.485\textwidth}
  \centering

  \includegraphics[
    width=\linewidth,
    keepaspectratio
  ]{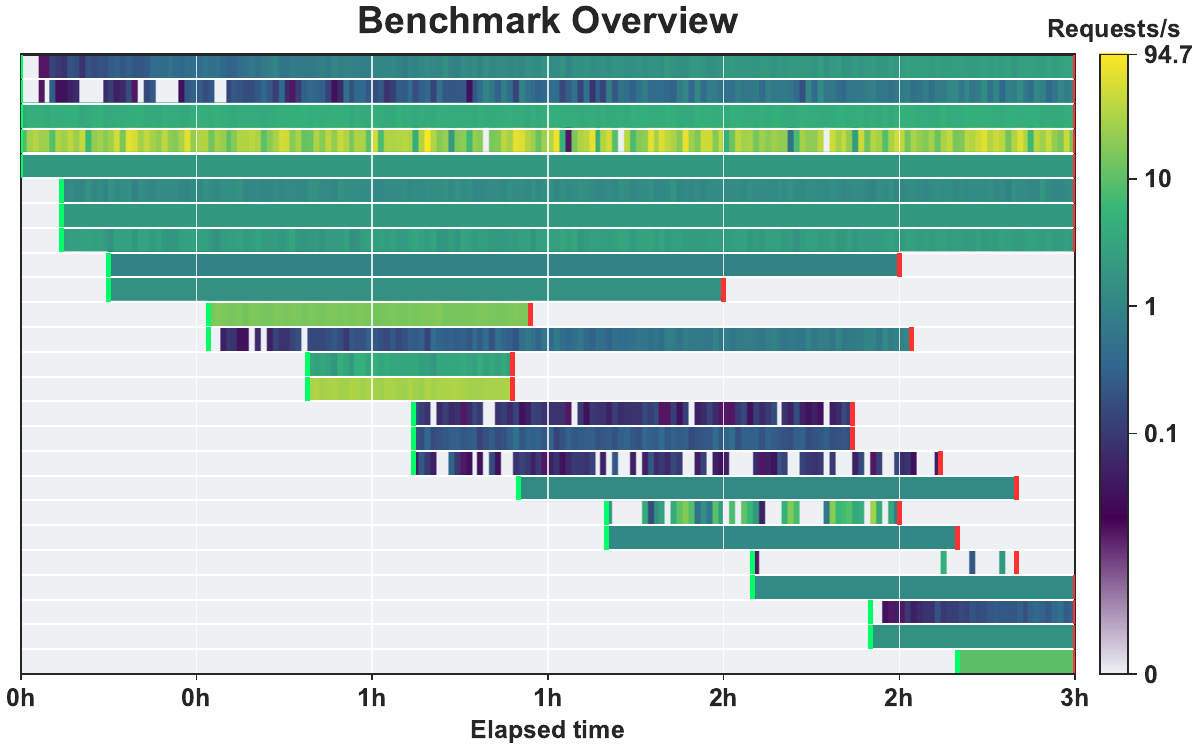}
    \captionsetup{justification=centering}
    \captionof{figure}{3-hour simulation benchmark schedule and traffic patterns.}
  \label{fig:benchmark-timeline}

  \Description{Timeline of jobs in the 24-hour heterogeneous benchmark.
  Online-job color intensity represents requests per second, while green and
  red markers identify scheduled admissions and ending points.}
\end{minipage}
\hfill
\begin{minipage}[t]{0.485\textwidth}
  \centering
  \includegraphics[
    width=\linewidth,
    keepaspectratio
  ]{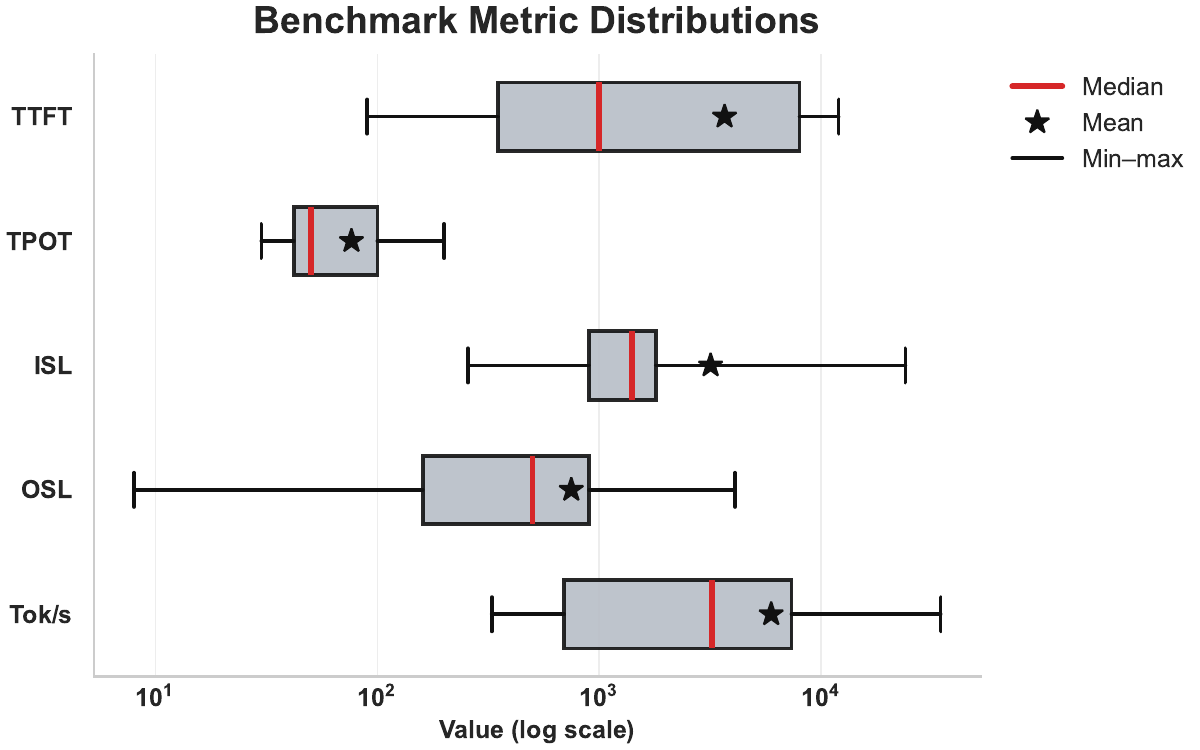}

    \captionsetup{justification=centering}
    \captionof{figure}{3-hour simulation benchmark workload distribution characteristics.}
  \label{fig:benchmark-characteristics}

  \Description{Horizontal box plots comparing configured time-to-first-token,
  time-per-output-token, input-sequence-length, output-sequence-length, and
  offered-token-rate values across benchmark jobs.}
\end{minipage}

\begin{minipage}{\textwidth}
\centering

\scriptsize
\setlength{\tabcolsep}{2.5pt}
\renewcommand{\arraystretch}{0.95}

\begin{tabularx}{\textwidth}{
  @{}
  >{\centering\arraybackslash}p{0.20\textwidth}
  >{\centering\arraybackslash}p{0.21\textwidth}
  >{\centering\arraybackslash}p{0.07\textwidth}
  >{\centering\arraybackslash}p{0.065\textwidth}
  >{\centering\arraybackslash}p{0.065\textwidth}
  C
  @{}
}
\toprule
Workload & Model & Mode & ISL & OSL & SLO / deadline \\
\midrule

Free multiturn chat
  & Llama-3.2-1B
  & Online & 900 & 300
  & TTFT $<$700 ms; TPOT $<$55 ms \\

Premium chat
  & Llama-3.3-70B
  & Online & 1,200 & 500
  & TTFT $<$600 ms; TPOT $<$38 ms \\

Code autocomplete
  & Qwen2.5-Coder-32B
  & Online & 1,800 & 120
  & TTFT $<$180 ms; TPOT $<$45 ms \\

Safety moderation
  & Llama-3.2-1B
  & Online & 256 & 8
  & TTFT $<$100 ms; TPOT $<$30 ms \\

Small-model chat
  & Llama-3.2-1B
  & Online & 256 & 64
  & TTFT $<$300 ms; TPOT $<$40 ms \\

Phi-4 long-context chat
  & Phi-4
  & Online & 9,000 & 220
  & TTFT $<$1.5 s; TPOT $<$55 ms \\

Small-model chat
  & Llama-3.2-1B
  & Online & 256 & 64
  & TTFT $<$300 ms; TPOT $<$40 ms \\

Support assistant
  & Mixtral-8x7B
  & Online & 1,400 & 350
  & TTFT $<$700 ms; TPOT $<$50 ms \\

Nightly evaluation
  & Llama-3.3-70B
  & Batch & 1,800 & 1,200
  & Completion deadline: 135 min \\

Phi-4 quantization probe
  & Phi-4
  & Batch & 1,000 & 512
  & Completion deadline: 105 min \\

Code morning burst
  & Qwen2.5-Coder-32B
  & Online & 2,200 & 96
  & TTFT $<$160 ms; TPOT $<$45 ms \\

Enterprise chat
  & Phi-4
  & Online & 1,500 & 420
  & TTFT $<$500 ms; TPOT $<$42 ms \\

Product-launch chat burst
  & Mixtral-8x7B
  & Online & 850 & 280
  & TTFT $<$650 ms; TPOT $<$50 ms \\

Moderation launch burst
  & Llama-3.2-1B
  & Online & 320 & 8
  & TTFT $<$90 ms; TPOT $<$30 ms \\

Agentic decode-heavy reasoning
  & Kimi-K2-Instruct
  & Online & 1,400 & 4,096
  & TTFT $<$2.5 s; TPOT $<$45 ms \\

GLM frontier reasoning
  & GLM-4.5-Air
  & Online & 1,800 & 2,500
  & TTFT $<$2.2 s; TPOT $<$50 ms \\

Small-active MoE reasoning
  & Qwen3-Next-80B
  & Online & 1,200 & 900
  & TTFT $<$1 s; TPOT $<$45 ms \\

Phi-4 calibration probe
  & Phi-4
  & Batch & 1,400 & 512
  & Completion deadline: 85 min \\

P0 safety replay
  & Llama-3.3-70B
  & Batch & 2,200 & 700
  & Completion deadline: 60 min \\

P0 safety reasoner
  & Phi-4
  & Online & 900 & 160
  & TTFT $<$350 ms; TPOT $<$40 ms \\

Long-context meeting summary
  & Qwen3-Next-80B
  & Online & 24,000 & 1,200
  & TTFT $<$3 s; TPOT $<$65 ms \\

Sales-report generation
  & Qwen3-Next-80B
  & Batch & 18,000 & 1,400
  & Completion deadline: 55 min \\

Legal review
  & Llama-3.3-70B
  & Online & 3,000 & 700
  & TTFT $<$1 s; TPOT $<$45 ms \\

Held-out frontier evaluation
  & DeepSeek-V3
  & Batch & 1,800 & 1,600
  & Completion deadline: 35 min \\

MoE exploration probe
  & Qwen3-Next-80B
  & Batch & 1,000 & 800
  & Completion deadline: 20 min \\

\bottomrule
\end{tabularx}

\captionsetup{justification=centering}
\captionof{table}{Workload overview of 3-hour simulation benchmark.}
\label{fig:benchmark_workloads}

\end{minipage}
\end{figure*}

\section{Deployment Configuration Space}
\label{app:knobs}

Table~\ref{tab:knobs} groups the configuration knobs that a chain sets into families, with representative knobs for each.

\begin{table}[h]
\centering
\small
\caption{Configuration knobs per chain, grouped by family.}
\label{tab:knobs}
\begin{tabular}{@{}l r p{0.44\columnwidth}@{}}
\toprule
Family & Knobs & Representative knobs \\
\midrule
Environment & \placeholder{?} & market, cloud, region, availability zone, GPU type, instance type, nodes per chain, interconnect \\
Parallelism & \placeholder{?} & TP, PP, DP, EP, SP, CP \\
Engine and runtime & \placeholder{?} & engine and version, runtime image, attention backend, CUDA graphs, torch.compile \\
Batching and scheduling & \placeholder{?} & max sequences, max batched tokens, max model length, chunked prefill and chunk size, scheduling, preemption, and router policies \\
Memory and KV cache & \placeholder{?} & GPU memory utilization, block size, KV-cache dtype, prefix caching, KV offloading, KV transfer method \\
Quantization & \placeholder{?} & weight and activation dtypes, weight quantization method and bits, activation quantization method \\
Speculative decoding & \placeholder{?} & method, draft model, speculative tokens, acceptance threshold \\
Disaggregation & \placeholder{?} & prefill/decode split, prefill and decode worker counts \\
\midrule
Total & $\sim$140 & \\
\bottomrule
\end{tabular}
\end{table}

\section{Baseline Implementation Details}
\label{app:baseline-implementation}

This section describes the non-\system baselines that were implemented. It excludes \system and its ablations.

\subsection{Common Store-to-plan contract}

Each implemented driver independently reads one frozen Store snapshot at a
planning tick: the resource map and the running, waiting, paused, and finished
job groups for the evaluated user.  It subtracts the footprint of running jobs
from the resource map before making new assignments.  It then emits exactly one
lifecycle action for every visible job: \texttt{keep} for running jobs,
\texttt{place} or \texttt{defer} for waiting jobs, \texttt{defer} for paused
jobs, and \texttt{preempt} for finished jobs.  A placement is serialized as a
normal Store plan and is checked against the concrete environment or pool
capacity before it is recorded.

CP-SAT, contextual Thompson sampling, Prism, and EvoX each locally construct
their candidate set from the same kinds of public facts, rather than reading
\system state.  For every waiting job, they read the model, token and SLO
fields, priority, and an explicit or benchmark VRAM requirement.  They enumerate
available resource targets and policy-permitted TP values, choose the smallest
PP sufficient for the VRAM requirement with DP fixed to one, and record each
candidate's resource keys, GPU footprint, price, model-weight estimate, and
KV-demand estimate.  The drivers reject candidates that violate the
configuration policy or nested environment/pool capacity.  This gives the four
methods the same job and resource-map interface while leaving their selection
rules independent of \system's predictions, mechanisms, and objective.

The contextual Thompson-sampling and EvoX drivers are currently registered in
the simulator CLI.  The best-fit, CP-SAT, AIConfigurator, and Prism drivers are
implemented and unit-tested Store adapters, but still need factory registration
before they can be selected through the CLI.

\subsection{Implemented baseline policies}

\paragraph{Best fixed configuration.}
The static control uses a checked-in mapping from benchmark job identity to
hardware target and TP/PP/DP configuration.  For each waiting job, it resolves
that mapping against the current resource map and greedily places the job only
when the required target still has capacity.  If a mapping is absent or no
longer fits, the implementation uses a VRAM-feasible fallback; otherwise it
defers the job.  The fallback lets the control produce a valid Store plan when a
named benchmark configuration cannot be bound to the current fleet.  It must be
reported as a fallback, not as a development-selected fixed configuration.

\paragraph{Rolling CP-SAT.}
CP-SAT receives the static candidates and capacity constraints above.  It creates
one Boolean variable for each job--candidate pair, constrains each job to at
most one selected candidate, and constrains every environment and pool footprint
to its available capacity.  The solver first maximizes priority-weighted job
admission.  With that value fixed, it minimizes hourly price, GPU footprint, and
a stable candidate-order tie break.  It uses one worker, seed zero, and a
declared time limit.  We use a rolling solve because capacity and the waiting
queue change each tick; the re-solve is necessary to make the static optimizer
operate on the actual current resource map rather than an offline snapshot.

\paragraph{Contextual Thompson sampling.}
The contextual-bandit baseline uses a Bayesian linear reward model over 11
public action features: an intercept; job priority and queue depth; average
input and output lengths; SLO tightness; candidate GPU count, memory, KV-cache
pressure, price, and requested capacity share.  Jobs are processed in priority
order.  For each job, the planner samples a weight vector from its current
linear-Gaussian posterior, scores the feasible candidates with that sample, and
chooses the highest-scoring candidate before reserving its capacity.  After the
simulator returns an outcome, the driver applies the corresponding rank-one
posterior update.  Online reward is the fraction of offered output tokens that
meet both latency targets, minus a normalized price penalty; batch reward is
zero when the deadline is missed.

We use Thompson sampling rather than LinUCB, epsilon-greedy, a neural bandit,
or a tree/forest bandit because the action set is discrete, the context is a
small explicit feature vector, and the controller needs to act from few
observations.  Sampling from the posterior performs probability matching: a
candidate is explored when uncertainty gives it a plausible chance of being
best, instead of using a fixed random-action rate or a separately tuned UCB
bonus.  The implementation is deliberately a linear Thompson sampler, not
Bayesian optimization and not a claim that it dominates the other bandit
families on these workloads.  Its default posterior-draw scale is 0.25 and its
random seed is zero for reproducibility.



\paragraph{Prism port.}
The Prism adapter translates a memory-pressure placement rule to the aggregate
resource map.  It orders jobs by decreasing maximum estimated KV demand and
then priority.  For each job, it filters out candidates that do not leave room
for the model weights or exceed remaining nested capacity, then chooses the
target with the least accumulated KV demand divided by residual memory.  Price,
GPU footprint, and candidate ID are deterministic tie breaks.  After selection,
the target's residual memory and KV-demand state are updated before placing the
next job.

This preserves the memory-centric placement idea while using Store-expressible,
exclusive shapes and the common safety checks.  It is a documented heuristic
port, not a claim that the full Prism artifact, including its complete sharing
and scheduling runtime, is executed unchanged.

\paragraph{Rolling EvoX.}
At each tick, EvoX receives the static candidate payload and nested capacities.
SkyDiscover evolves a deterministic Python function,
\texttt{compute\_placement(payload)}, that returns one candidate ID or a defer
decision for every waiting job.  The evaluator rejects malformed or infeasible
programs and scores valid outputs by priority-weighted admission, followed by
hourly price and GPU-footprint tie breaks.  The initial program is a simple
priority-first greedy policy; the best program from one tick seeds the next
tick's search.  The winning program is rerun on the frozen payload, its choices
are revalidated, and valid choices become Store placement actions.

EvoX is the search procedure in this baseline; there is no second optimizer.
Candidate construction and feasibility checks stay in the simulator so evolved
code cannot invent hardware or bypass capacity rules.  The rolling driver adapts
to the current queue and resource map, but its EvoX evaluator currently uses
static placement facts rather than simulated serving outcomes as its fitness
signal.  The frozen-snapshot EvoX table uses the same candidate encoding and
evaluator without the cross-tick lifecycle loop.

\subsection{Real hardware configuration}

Each of the real experiments are conducted on 3 Google Kubernetes Engine (GKE) clusters, in the following regions: us-central1 (Iowa, USA), us-east5 (Ohio, USA), europe-west2 (London, UK). The 3 clusters combine to a fleet of 104 heterogeneous GPUs, consisting of a mix of Nvidia H100-80GB, RTX PRO 6000, A100-80GB and A100-40GB. The H100s and A100s are placed in 8-GPU instances, and the RTX PRO 6000 are placed in both 8-GPU instances and 4-GPU instances.

\clearpage

\end{document}